\documentclass[useAMS,usenatbib,usedcolumn]{mn2e}
\usepackage{multirow}
\usepackage{amsmath}
\usepackage{graphicx,epsfig}
\usepackage[graphicx]{realboxes}
\usepackage{float}

\title[Open Clusters' Membership Probabilities]{Stellar Open Clusters' Membership Probabilities: an N-Dimensional Geometrical Approach.}

\author[Laura Sampedro and Emilio J. Alfaro]
{Laura Sampedro and Emilio J. Alfaro
\\IAA-CSIC, Glorieta de la astronom\'ia S/N. 18008, Granada. Spain
\\sampedro@iaa.es, emilio@iaa.es}

\date{Submitted to MNRAS}
\date{Accepted XXX. Received YYY; in original form ZZZ}

\pubyear{2015}

\begin{document}
\label{firstpage}

\pagerange{\pageref{firstpage}--\pageref{lastpage}} 
\maketitle{}

\begin{abstract}

We present a new geometrical method aimed at determining the members of open clusters. The methodology estimates, in an N-dimensional space, the membership probabilities by means of the distances between every star and the cluster central overdensity. It can handle different sets of variables, which have to satisfy the simple condition of being more densely distributed for the cluster members than for the field stars (as positions, proper motions, radial velocities and/or parallaxes are). Unlike other existing techniques, this fact makes the method more flexible and so can be easily applied to different datasets. To quantify how the method identifies the cluster members, we design series of realistic simulations recreating sky regions in both position and proper motion subspaces populated by clusters and field stars. The results, using different simulated datasets (N = 1, 2 and 4 variables), show that the method properly recovers a very high fraction of simulated cluster members, with a low number of misclassified stars. To compare the goodness of our methodology, we also run other existing algorithms on the same simulated data. The results show that our method has a similar or even better performance than the other techniques. We study the robustness of the new methodology from different subsamplings of the initial sample, showing a progressive deterioration of the capability of our method as the fraction of missing objects increases. Finally, we apply all the methodologies to the real cluster NGC 2682, indicating that our methodology is again in good agreement with preceding studies.

\end{abstract}


\section{Introduction}
\label{intro}

The determination of the members of a stellar open cluster is an essential prior task to a large number of astrophysical problems, mainly concerning the star formation process, the birth and destruction of stellar clusters, stellar evolution, Galactic structure and evolution, and many others. This task has a statistical nature and usually involves the separation of two populations defined by several variables of different natures: the cluster members and the field stars. To make a good classification, the quality, quantity and availability of the stellar variables are fundamental.  

Consequently, the advent of the new generation of large Galactic surveys like the Gaia-ESO Public Spectroscopic Survey (GES, \cite{2012Msngr.147...25G}; \cite{2013Msngr.154...47R}), or the Gaia mission \citep{2001A&A...369..339P}, will help us to enhance the current knowledge of the physics of our Galaxy and, in particular, of the Galactic star cluster populations. The Gaia mission will provide an unprecedented precision in astrometry that will result in very accurate measurements of positions, parallaxes and proper motions for one billion stars. This will enable us to build the first 5-D map of our Galaxy. If we add the Gaia radial velocities or the high precision ones given by GES, we will have a 6-D map covering the phase space for an important sample of the Milky Way stellar population. GES also intends to provide not only radial velocities but also chemical abundances and other sets of astrophysical parameters, increasing the number of phase-space dimensions to more than 12-D \citep{2012Msngr.147...25G}. 

Most of the current techniques address the estimation of the membership probabilities computing the probability density functions (hereafter pdfs) of the variables used in the analysis, either by parametric or non-parametric techniques. Examples of these approaches are: \cite{1958AJ.....63..387V} along with \cite{1971A&A....14..226S}, \cite{1985A&A...150..298C}, \cite{1990A&A...237...54Z}, \cite{1990A&A...235...94C} and \cite{2006SerAJ.173...57U}, among others. This fact makes the cluster membership determination dependent on the availability of these variables. But, being aware of all the possibilities that the new generation of galactic surveys offer, it becomes necessary to develop new tools that make it possible to exploit the sets of variables in a flexible way, being able to adjust the determination of the cluster members to the availability and characteristics of the data. 

The purpose of this study is, therefore, the determination of the potential members of a stellar cluster allowing for the use of as many phase-space variables (positions, parallaxes, proper motions, radial velocities) as possible. To achieve this objective, we have designed a geometrical approach, based on the distance distribution between each star of the sample and a central overdensity in an N-dimensional space (hereafter, N-D space). We assume that the distance distribution can be approximated by a mixture of two 1-D Gaussian functions: one for the cluster members and another for the field stars. In this way our method computes the pdfs of just one variable: the distance defined in an N-D space. Thus we can choose the variables in our study, being able to address the determination of cluster members from different perspectives. 

We tested the new methodology through its application to a series of realistic simulations of a sky region, in the position and proper motions subspaces, where a cluster and field stars co-exist. We compared our results with those obtained by applying two other techniques, one based on the parametric definition of the pdfs \citep{1985A&A...150..298C} and the other on the pdf direct kernel estimation \citep{1990A&A...235...94C}, always using the same simulated datasets. Finally, we selected the open cluster NGC 2682 to show the application of our approach to real data.

The paper is organised as follows: Section 2 details the fundamentals of the method; Section 3 describes the performed simulations and the figures of merit introduced to test it; the results achieved by the new methodology and the comparison with those obtained by other techniques are shown in Section 4, along with the results accomplished when we apply the methodology to the open cluster NGC 2682; and, finally, Section 5 discuss the results and highlights the main conclusions.

\section{METHODOLOGY}
\label{methodology}

Our approach considers an N-D space, where we estimate the distances between every star and a central overdensity. We assume that the N variables satisfy the basic condition of being more densely distributed for the cluster members than for the field stars. The distance distribution of cluster stars would therefore show a mean and a dispersion smaller and narrower than those for the field stars. A Bayesian analysis of distance pdfs will allow us to assign a cluster membership probability to the sample stars. In the following, we describe the foundations and application protocol of our methodology.

\subsection{Distances in an N-Dimensional space.}

Given the different nature of the variables involved in this problem, we need to normalise them. The normalisation is highly affected by the presence of outliers in the sample, for which the first step is the data pruning, removing those objects with a high probability of being outliers. In addition, the outlier detection represents a fundamental step in this procedure because these objects modify the estimated distribution functions both of the cluster and field stars, biasing the final membership probability.

For the pruning of outliers, we followed the OUTKER procedure proposed by \cite{1985A&A...150..298C} for the case of the proper motion distribution, and easily tied to the case of just one velocity dimension. It is important to note that an outlier is one of the few mathematical concepts that are not rigorously defined. In fact, we can never deterministically say that a given object is an outlier, above all if the only information that we have is the sample of objects. However, we can determine the probability of a given object from the sample being an outlier from the probability density function defined by the sample itself. The best approximation to the definition of an outlier is that of an object that is located in a low-density-probability region of the space of N-variables. For its detection, the OUTKER procedure compares the probability density observed for each object with the probability density function expected for the whole sample. Thus we obtain the probability of being an outlier for every object in the sample. Given these probabilities we can decide which objects to remove from the initial sample. As in the rest of this work we have followed the \textit{Bayes minimum error rate decision rule} \citep{bayes_rule}, which specifies a threshold in the membership probability of 0.5 to minimise the probability of error in the classification. Thus, those objects with a probability of being an outlier greater than 0.5 were removed from this analysis. This process is carried out just once at the beginning of the procedure and before the normalisation of the variables.

In principle one should detect and remove the outliers for all the variables involved in the problem, but due to the fact that the spatial distribution of the open clusters shows a highly variable case history of alternating regions of high and low density of objects, we have taken the decision to remove the outliers in the proper motion distributions.

\begin{figure*}
\includegraphics[width=15cm]{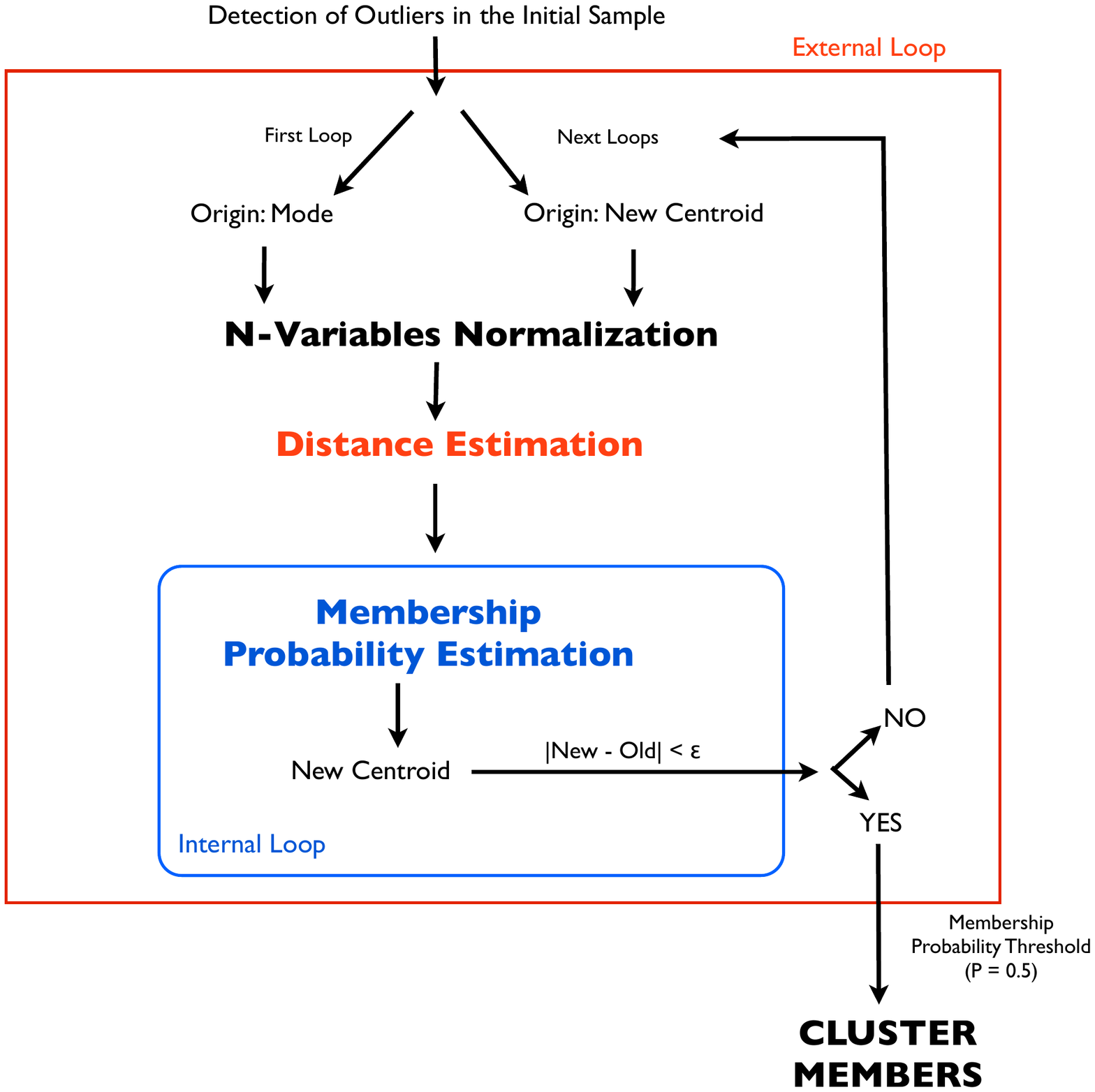} 
\caption{Flow diagram of the entire process.}
\label{ProcessScheme}
\end{figure*}

With an outlier-free sample, the estimation of the membership probabilities involves two iterative process, one inside the other. Figure \ref{ProcessScheme} details the flow diagram of the process. The external loop computes the distances, while the internal one deals with the membership probabilities estimation. During the first iteration, the variables used in the membership analysis (positions and/or proper motions and/or radial velocities...) are normalised by their modes and dispersions, according to: 

\begin{equation}
   X_i = \frac{x_i - x_0}{\sigma_x}
\end{equation} 

\noindent where $X_i$ represents the normalised variables and $x_i$ the initial ones for the $\textit{i}$-th star, $x_0$ is the mode of the $x_i$ distribution and $\sigma_x$ its standard deviation. Then the distances for every star \textit{$(dis_i)$} are computed making use of the expression:

\begin{equation}
   \textit{$dis_i$} = \sqrt{(X_{i_N})^{T}(M)(X_{i_N})}
\end{equation} 

\noindent where \textit{$(X_{i_N})$} and \textit{$(X_{i_N})^{T}$} are the N-dimensional vector and the transpose vector respectively, composed by the values taken from the normalised N-variables for the \textit{i}-th star. In the present work, we adopt an \textit{euclidean metric}, \textit{M}. However, it could be changed, allowing the variables to have different weights more according to the characteristics of the problem.

\subsection{Bayesian Membership Probabilities.}

These distances are used to estimate the membership probabilities supposing that the distance distribution can be fitted by a mixture of two 1-D Gaussians: one for the cluster members and another for the field stars. Considering that the subscripts \textit{c} and \textit{f} refer to the cluster members and to the field stars, respectively, the pdf model of the distance distribution is given by: 

\begin{equation}
   \phi_i(dis_i) = n_c \phi_{i,c}(dis_i)+ n_f \phi_{i,f}(dis_i)
\end{equation}

\noindent where \textit{$n_c$} and \textit{$n_f$} are the priors, and \textit{$\phi_{i,c}(dis_i)$} and \textit{$\phi_{i,f}(dis_i)$} are the conditional pdfs defined as follows:

\begin{equation}
   \textit{$\phi_{i,c}(dis_i)$} = \frac{1}{\sigma_c \sqrt{2\pi}}
   exp\left(-\frac{1}{2} \left(\frac{dis_i-\mu_c}{\sigma_c} \right)^2 \right) 
\end{equation}

\begin{equation}
   \textit{$\phi_{i,f}(dis_i)$} = \frac{1}{\sigma_f \sqrt{2\pi}}
   exp\left(-\frac{1}{2} \left(\frac{dis_i-\mu_f}{\sigma_f} \right)^2 \right) 
\end{equation}

\noindent where \textit{$dis_i$} is the distance value for the \textit{i}-th star and $\mu_c$, $\sigma_c$, $\mu_f$ and $\sigma_f$ are the Gaussian model parameters of both populations. 

Through an iterative Wolfe estimation procedure \citep{Wolfe} and starting with some reasonable values for the Gaussian model parameters (means, standard deviations and priors of both groups of stars), the algorithm computes the membership probabilities (within what we call the internal loop) according to:

\begin{equation}
   \textit{Prob$_i(c/dis_i)$} = \frac{n_c\phi_{i,c}(dis_i)}{\phi_i(dis_i)} 
\end{equation}

These probabilities are used to derive a new estimation of the model parameters that are then used to recompute the pdfs and to update the membership probabilities until the convergence of the iterative process is reached. This convergence is reached once the difference between the parameters that define the Gaussians of the distance distribution are similar between one iteration and another with a tolerance of one thousandth. 

The resulting membership probabilities are used to compute the weighted mean of every N variable used in the distance estimation in order to re-determine the cluster center in the N-D space. The new cluster center is compared with the previous one (in the case of the first iteration with the mode of the variables). If the difference is larger than a chosen threshold ($\epsilon$$>$0.001), another iteration is performed where the distances are now computed from the new centroids. In another case, the convergence is reached, and we get the final membership probabilities. 

For the range of values obtained in our simulations, the thresholds of 0.001 ensure that the differences in the cluster membership probabilities and in the distance distribution obtained between the last iteration and the one before, guarantee us the same classification of objects between cluster and field.

It should be noted that, although in the first iteration the distances are computed using the mode, i.e., the overdensity in the N-D space, in subsequent iterations the distances are calculated from the estimated cluster centroid. This way of proceeding is due to the fact that in the first iteration there is no previous classification of the stars into the two categories, and therefore we do not know the centroid of the cluster. As a first hypothesis we consider that the distances to any point of origin must have the histogram maximum in the cluster centroid. If, due to the characteristics of the sample, this maximum corresponds to the field stars, the situation is corrected in the following iterations where it is a requirement (as starting hypothesis) for the dispersion of the cluster stars to be lower than that of the field stars. Once the first probabilities distribution has been obtained, we obtain the cluster centroid as the moment of zero order of this distribution, which thus becomes the new origin for the determination of distances of the next iteration.

\begin{figure}

\begin{center}
\includegraphics[width=9cm]{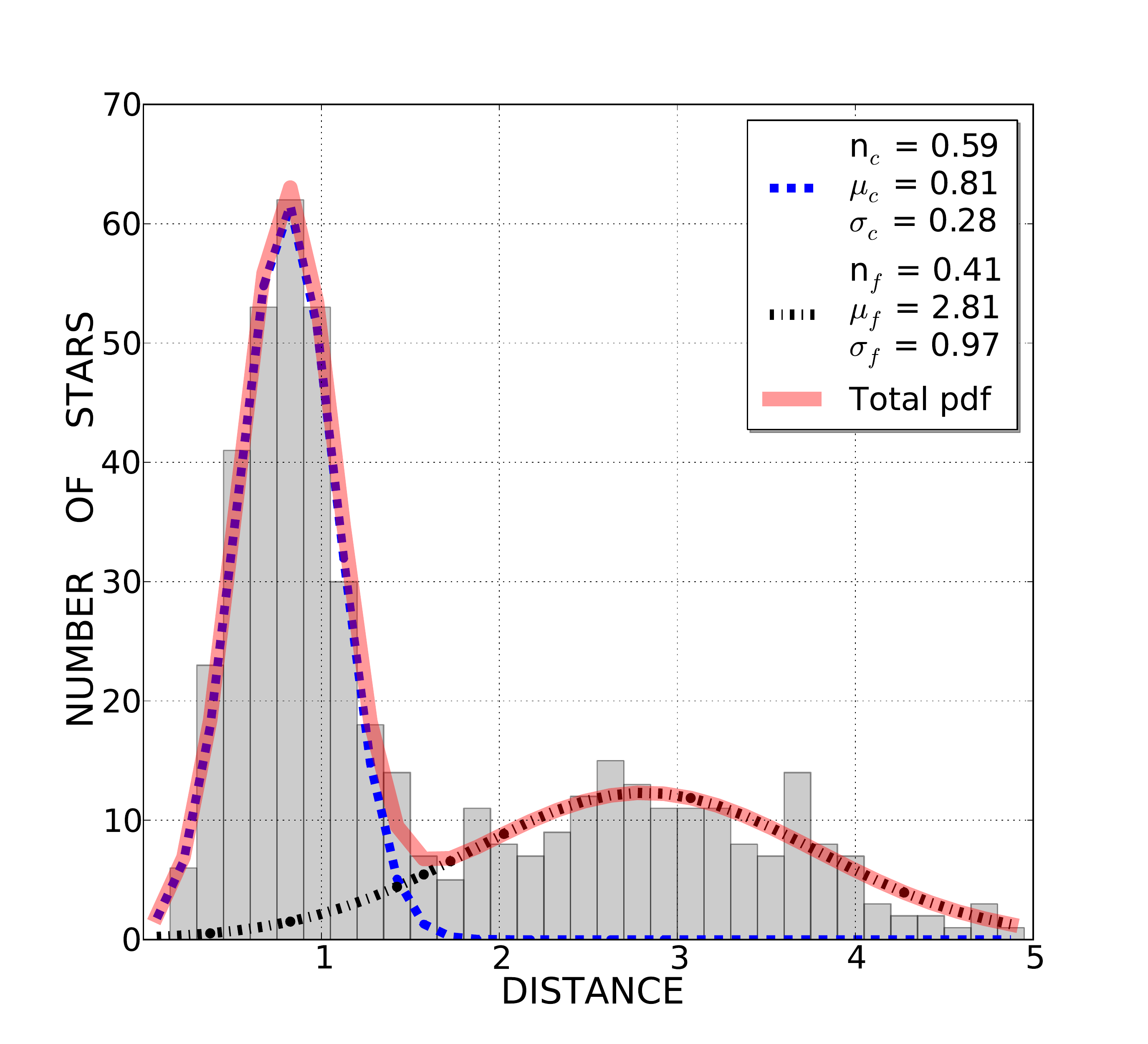} 
\caption{Example of distance distribution, computed using the simulated positions and proper motions data. The legends show the best fit parameters from which the fit to two 1-D Gaussians is done. The dashed blue and the dot-dashed black lines refer to the cluster and field distance distributions. The total pdf is over-plotted as a solid red line.}
\label{dis1}
\end{center}

\end{figure}

Figure $\ref{dis1}$ shows an example of a distance distribution and the best fit to two 1-D Gaussians, in which the positions and proper motions are the variables used for the cluster membership determination (see the following section for the description of the simulations). The dashed blue line represents the Gaussian fit for the cluster distance distribution, while the dot-dash black line represents the Gaussian fit for the field stars. The total pdf is over-plotted as a solid red line. As can be seen in the figure, the Gaussian fit to the cluster distance distribution shows a lower mean and a narrower dispersion than the field stars. This is an expected result, since the distances were computed with respect to the cluster center. In addition, it is the narrowest distribution due to the fact that we are using variables which are more densely distributed for the cluster members than for the field stars.

Once our methodology estimates the membership probabilities, it is necessary to make a decision on the probability value from which a star will be classified as a cluster member. Again we used the decision criterion of the \textit{Bayes minimum error rate decision rule}. Thus, any star with a cluster membership probability above 0.5 will be classified as a cluster member, so providing the final classification of the sample into the two populations.

\section{Simulations}
\label{simu}

\subsection{Simulated Cases}

In order to test the potential of the methodology to separate between cluster members and field stars, we designed a series of simulations, recreating regions of the sky including a stellar open cluster and field star distribution. The simulations were performed in the position and in the proper motion subspaces with different sets of parameters chosen to quantify the feasibility of this method for determining the final star classification into both populations. Trying to make the simulations as realistic as possible, parameters for both population distributions were selected, taking into account the current data in the main and most complete stellar open cluster catalogues. In particular, we make use of the values listed in the work of \cite{2014A&A...564A..79D}, where a sample of 1805 clusters was compiled and analysed. The distances at which the clusters are found have been taken from the DAML02 catalogue \citep{2002A&A...389..871D}. Making use of these two studies and imposing the condition that the number of objects in the cluster field be lower than 5000, we obtain the parameters of the cluster and field proper motions distributions, the distances and the projected radii for a sample of 1646 clusters.

In order to simulate the sample, we have calculated the average of the number of objects in the field of the 1646 clusters, obtaining a value that we approximate to 500 stars. Given the high degree of skewness of the distance distributions and the clustersÕ radii, we decided to utilise their respective modes as a representative value. Thus we consider that a typical cluster, in this catalogue, is found at a distance of 1250 pc and has a radius of 2 pc.

The spatial centroid of the stars of the cluster is always located in the central position of the sample. Considering that the real radius of the cluster is 2 pc and it is found at a distance of 1250 pc, we obtain a projected radius value of 0.093$^\circ$. The distribution in position of the cluster stars is given by a circular Gaussian function with constant dispersion for all cases and equal to ($\sigma_{pos}$ =) 0.031$^\circ$, which is equal to a third of the projected radius. The spatial distribution of the field stars is an homogeneous distribution within a square field with a side of approximately 0.3$^\circ$.

Both the proper motions of the cluster and the field stars are defined by a bivariate Gaussian pdf, circular for the cluster and elliptic for the field. The field's proper motions distribution is always centered on the (0,0) and its covariance matrix is diagonal with values that vary (10, 15, 20 and 25) mas/yr for $\sigma_{\mu f,x}$, and 1 or 1.2 times $\sigma_{\mu f,x}$ for the dispersion on the y axis ($\sigma_{\mu f,y}$). These values correspond to the most frequent interval in the distribution of the quotient $\sigma_{\mu_{f,x}}$/$\sigma_{\mu_{f,y}}$, for the distribution of the field stars for the sample of 1646 clusters.

The relative frequencies of cluster stars used are 20\%, 40\%, 60\% and 80\% of the total sample. The internal dispersion of the cluster's proper motions is constant for all cases and has been determined as the mean of the velocity in virial equilibrium for all of the values of the fraction of cluster stars, considering a mass equal to the number of stars in the cluster and a radius of 2 pc, which gives us a value of 0.42 km/s per degree of freedom. Considering that the cluster is at a typical distance of 1250 pc, a value of 0.07 mas/yr for the internal dispersion of the proper motions of the cluster stars is obtained.

However, the proper motions distribution observed in a cluster is mainly dominated by the observational errors. Once again utilising the values catalogued by \cite{2014A&A...564A..79D}, we can see that the standard deviation of the proper motions of the clusters ($\sigma_{\mu c}$) presents a maximum around 3 mas/yr, a value that we have taken to model the proper motions distribution of the cluster in our simulations. In order to introduce the errors into the sample, the initial proper motion for each star (obtained using the internal velocity dispersion) has been replaced by a random number taken from a Gaussian distribution of equal mean to the initial proper motion and with standard deviation equal to the error, which in this case is 3 mas/yr. The position of the cluster in the Vector Point Diagram (VPD) keeps its component $\mu_{c,x}$ fixed to 0 mas/yr and the component $\mu_{c,y}$ varies between the following values: 1, 5 and 7.5 mas/yr.  In the end, a total of 96 simulations have been constructed to quantify the new methodologyÕs potential.

\begin{figure*}
\begin{center}
\includegraphics[width=14.cm]{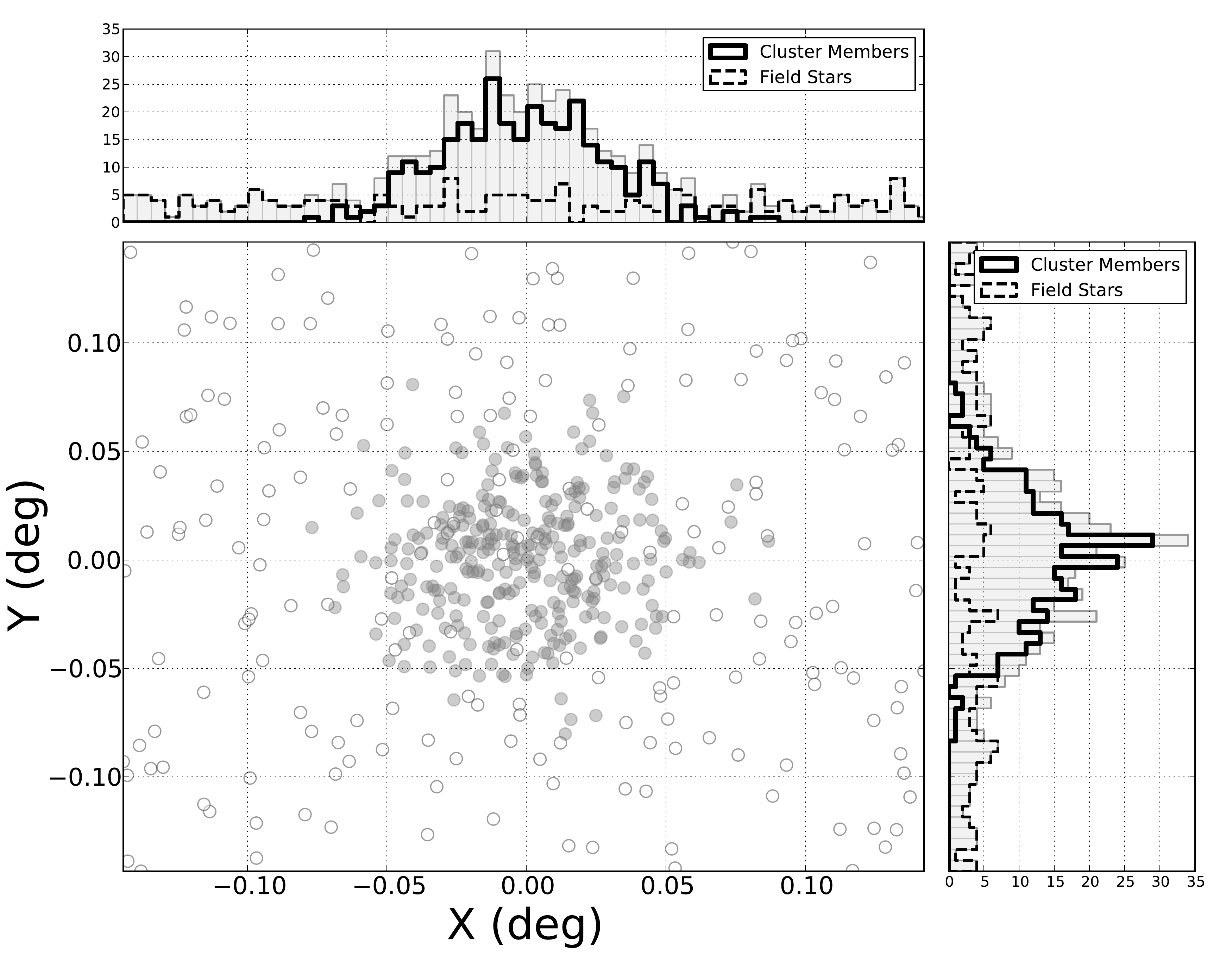} 
\includegraphics[width=14.cm]{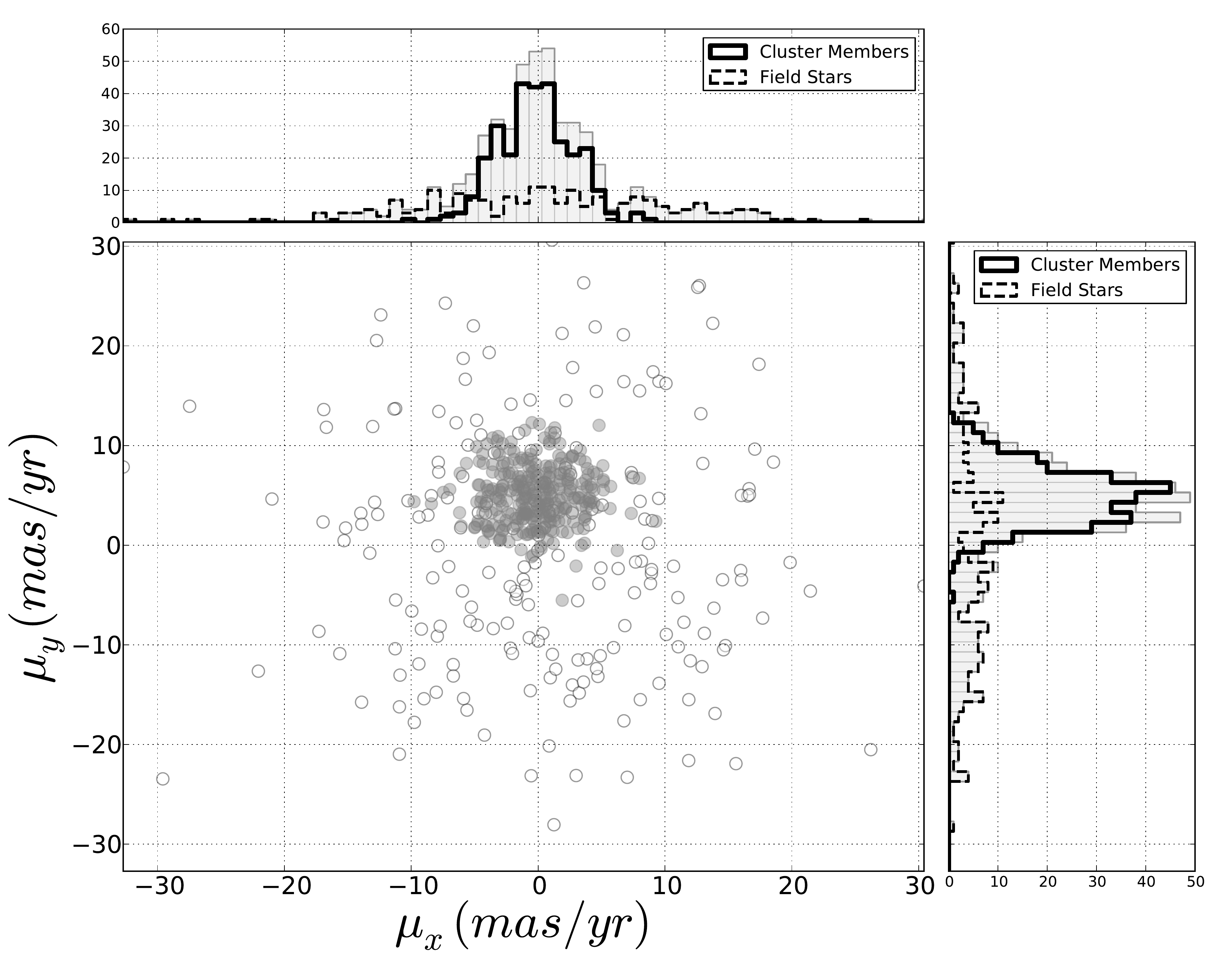} 
\caption{An example of distributions of a simulated cluster and field stars in the positions subspace at the top, and in the proper motions subspace at the bottom. The cluster members are distributed following a circular Gaussian distribution in both subspaces. The field stars follow a random distribution in positions and an elliptical Gaussian distribution in the proper motion subspace. This configuration corresponds with a percentage of 60 \% of cluster members, $\mu_{c,y}$ = 5.0, $\sigma_{\mu f, x}$ = 10 mas/yr and $\sigma_{\mu f,y}$ = 12 mas/yr.}
\label{sim}
\end{center}
\end{figure*}

In Tables \ref{table1} and \ref{table2} a summary is shown of the set of parameters utilised in the simulations for the four variables of the sub-phase space. Table \ref{table1} summarises the fixed parameters of the simulations and Table \ref{table2} the variables. An example of the simulations carried out is shown in Figure \ref{sim}, where the upper and lower graphs show the distribution of the field and cluster stars in the positions and proper motions subspaces, respectively.

With the objective of analysing the possible bias introduced by an observational subsampling of the population, we have obtained a subsample for each simulation of 50\%, 20\% and 10\% with regard to the initial number of objects simulated (250, 100 and 50 objects, respectively), chosen randomly though keeping the proportions between field and cluster stars. 

With these four variables we have carried out different experiments utilising subspaces of phase space of N = 1, 2 and 4 dimensions. In the case of 1 and 2 dimensions we restricted ourselves to the kinematic data. All the simulations have been tailored using the pdfs, for the different statistical distributions, given by NumPy Package in Python.

\begin{table*}
\caption{{\small Fixed parameters in all the simulations.}}
\label{table1}
\scalebox{1.1}[1.1]{
\begin{tabular}{|l|l|}
     \hline
     \hline
	 \textbf{Parameters}                                                                 & 	\textbf{Values}    \\
	 \hline
     \hline
	 Total Number of the Star Sample                                         &       N$_{Total}$ = 500	 \\
 	 \hline
	 Cluster and Field Centroids in Space                                  &       $(x_c,y_c)$ = $(x_f,y_f)$ = (0,0) degree \\
 	 \hline
	 Cluster Proper Motion Centroid (X-coord)                         &       $\mu_{c,x}$ = 0 mas/yr \\
     \hline
	 Field Proper Motion Centroid                                               & 	   $(\mu_{f,x},\mu_{f,y})$ = (0,0) mas/yr \\
 	 \hline
	 Cluster Proper Motion Dispersion              & 	 $\sigma_{\mu c, x}$ = $\sigma_{\mu c, y}$ = $\sigma_{\mu c}$ = 3 mas/yr  \\
 	 \hline
	 Cluster Angular Radius                                                       &     $R_{Cluster}$ = 0.093 degrees	 \\
     \hline

\end{tabular}}
\end{table*}

\begin{table*}
\caption{{\small Variable parameters in the simulations.}}
\label{table2}
\scalebox{1.1}[1.1]{
\begin{tabular}{|l|l|}
     \hline
     \hline
	 \textbf{Parameters}                                                                            &  	 \textbf{Values}      \\
 	 \hline
     \hline
 	 Cluster Proper Motion Centroid (Y-coord)                                      & 	  $\mu_{c,y}$ = (1, 5, 7.5) mas/yr   \\
  	 \hline
  	 Proper Motion Dispersion of the Field Stars (X-coord)              & 	   $\sigma_{\mu f,x}$ = (10, 15, 20, 25) mas/yr  \\
  	 \hline
 	 Proper Motion Dispersion of the Field Stars (Y-coord) &     $\sigma_{\mu f,y}$ = (1, 1.2)$\cdot \sigma_{\mu f,x}$  \\
         \hline
         Fraction of Cluster Members                                                      &     20\%, 40\%, 60\%, 80\%       \\
         \hline
         Subsampling                                                                               &     50\%, 20\%, 10\%       \\
     \hline

\end{tabular}}
\end{table*}

\subsection{Figures of Merit.}

With the purpose of studying in detail the potentiality of the different methods that we analyse comparatively in this work, we have considered two figures of merit, \textit{Completeness (C)} and \textit{Misclassification (M)} that are defined by the following expressions:

\begin{equation}
   C = \frac{Nº_{c,met}}{N_{c,real}}
\end{equation} 

\begin{equation}
   M = \frac{Nº_{c->f,met} + Nº_{f->c,met}}{Nº_{Total}}
\end{equation} 

\noindent where Nº$_{c,met}$ is the number of simulated cluster members recovered by the different methodologies, N$_{c,real}$ is the total number of simulated cluster members, Nº$_{c->f,met}$ is the number of cluster members classified as field stars, Nº$_{f->c,met}$ is the number of field stars classified as cluster members, and Nº$_{Total}$ is the number of stars in the total sample.

\section{Results}
\label{intro}

In this section we analyse the results of applying the new methodology to the 96 models that have been simulated. For this we make use of the previously defined figures of merit. One of the aims of this work is to compare the new methodology with other already defined methods that have been widely used. These are the parametric method for the proper motion distributions (MT1 for the 1D case and MT2 for the 2D case), and the non-parametric method applicable to the case of the four dimensions of sub-phase space (hereafter MT4). 

For MT1 and MT2 we use the formalism introduced by \cite{1985A&A...150..298C} for proper motion distribution, which is easily reducible to the case of one dimension (see, for example, \cite{2014A&A...569A..17C}). This technique approaches the total pdf as a mixture of two bivariate Gaussian distributions (for 1 or 2 dimensions): one for the stellar cluster and another for the field population. Through an iterative Wolfe estimation method, the pdfs' parameters are determined as well as the corresponding membership probabilities. In all cases we used the \textit{Bayes minimum error rate decision rule} to make the final classification. 

For the case of 4 dimensions, we compare the results obtained from the membership analysis based on the non-parametric method (MT4) developed by \citep{1990A&A...235...94C}. This method doesn't make any a priori assumptions about the cluster and field star distributions and assumes two hypotheses: i) there are two populations, cluster members and field stars, and ii) the cluster members are more densely distributed than the field stars in any subspace of variables. Membership probabilities are calculated using Kernel estimators in an iterative way. In every iteration, 3 different probabilities for each star are estimated: one just using the positions of the stars, the second for the proper motion data (kinematic probability), and the last using both positions and proper motions (joint probability). Cluster members in every iteration are selected as those stars with joint and kinematic probabilities higher or equal to 0.5.

To summarise, MT1 is applied to the 96 simulations where the only variable used is $\mu_{y}$. MT2 is also applied to the 96 simulations, where the variables used are the proper motions ($\mu_{x}$, $\mu_{y}$), while MT4 is applied to the same number of test cases using both position and proper-motion variables (x, y, $\mu_{x}$, $\mu_{y}$).

It is clear that the goodness of the classifications into cluster and field stars depends on the characteristics of the distribution functions that define the two populations in the sub-phase space. In other words, it depends on the heteroscedasticity of the pdfs. A measure of this is given by the Chernoff distance (\cite{Chernoff}, hereafter CD), which is a measurement of the degree of similarity between the distribution functions that describe both populations. This CD is calculated making use of the parameters of the simulated distributions, both the positions and the proper motions, of both populations. Its general analytic expression is given by the equation:

\begin{equation}
\begin{split}
   CD & = \frac{1}{2} \alpha_c \alpha_f (\mu_c - \mu_f)^T[\alpha_c \Sigma_c + \alpha_f \Sigma_f]^{-1} (\mu_c - \mu_f) \\
            & + \frac{1}{2} log \frac{| \alpha_c \Sigma_c + \alpha_f \Sigma_f |}{|\Sigma_c|^{\alpha_c} |\Sigma_f|^{\alpha_f}}
\end{split}
\end{equation} 

\noindent where $\alpha$, $\mu$ and $\Sigma$ are the percentages, means and covariances of both groups of stars, and the superscripts T and -1 refer to the transpose vector and to the inverse of the matrix, respectively.

The methodology proposed in this work will also be analysed according to different observational subsamples of the same distribution function of the phase space. This means that, using the earlier simulations, we will extract samples corresponding to 50\%, 20\% and 10\% of each of them, in all cases keeping the same proportion of cluster stars and field stars as in the original test case.   

In the next part we will analyse the behaviour of the different methodologies for the simulated samples, as well as for cluster NGC 2682.

\subsection{Analysis of the Simulations.}

In this work we present a new methodology in which the implementation of the outlier determination must be the first step in the membership analysis. In the simulations carried out we have assumed some well-behaved errors (Gaussian errors with zero mean and standard deviation of 3 mas/yr). For these simulations, the purging or not of outliers leads to variations of less than 1\% for the completeness (\textit{C}) and 0.4\% for the misclassification rate (\textit{M}). However, this could be different for the actual samples where the errors might be larger and their distribution not necessarily well behaved. No simulated cluster member has been removed as an outlier in our simulations. 

\begin{figure*}
\begin{center}
\includegraphics[width=7.4cm]{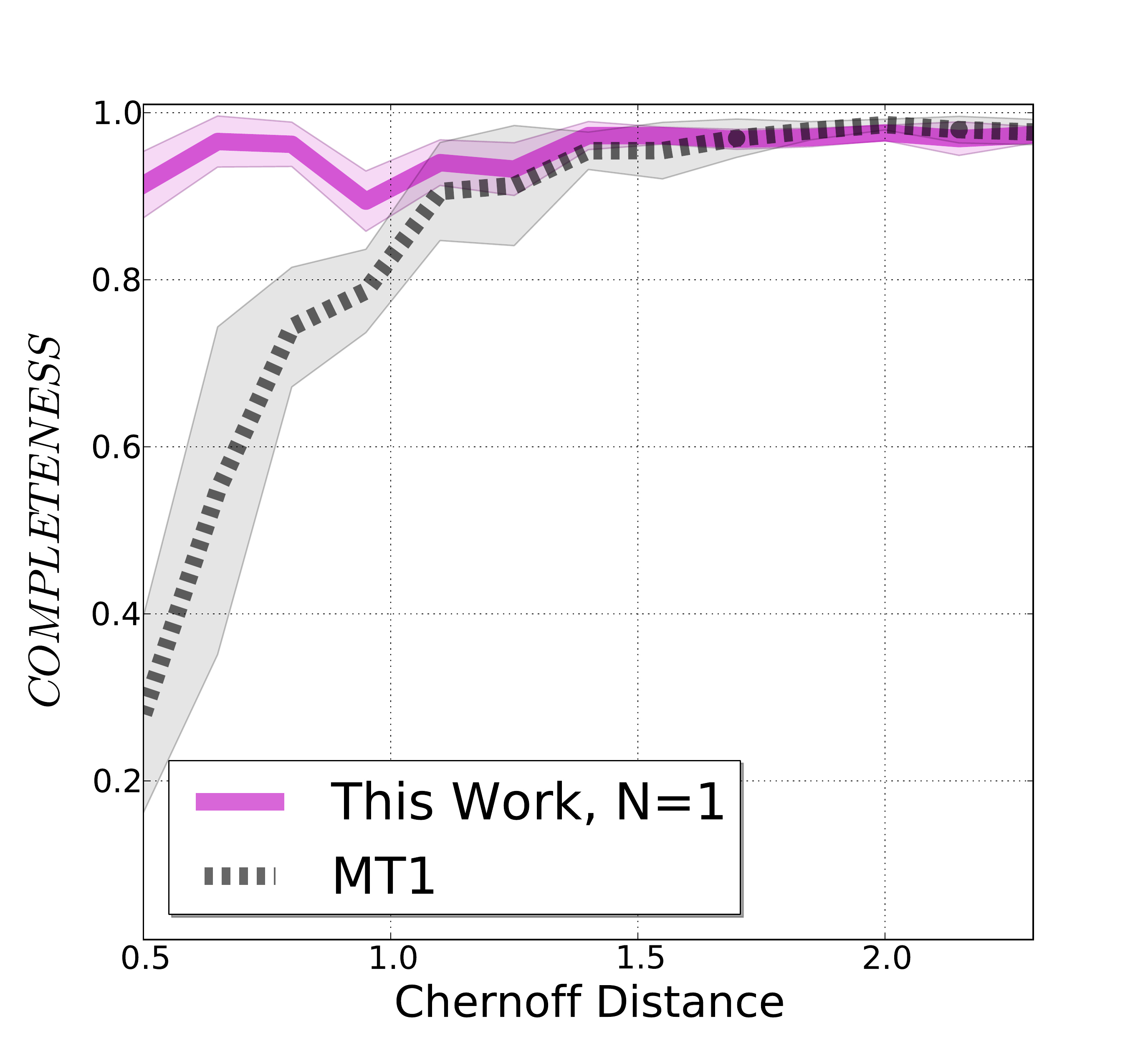} 
\includegraphics[width=7.4cm]{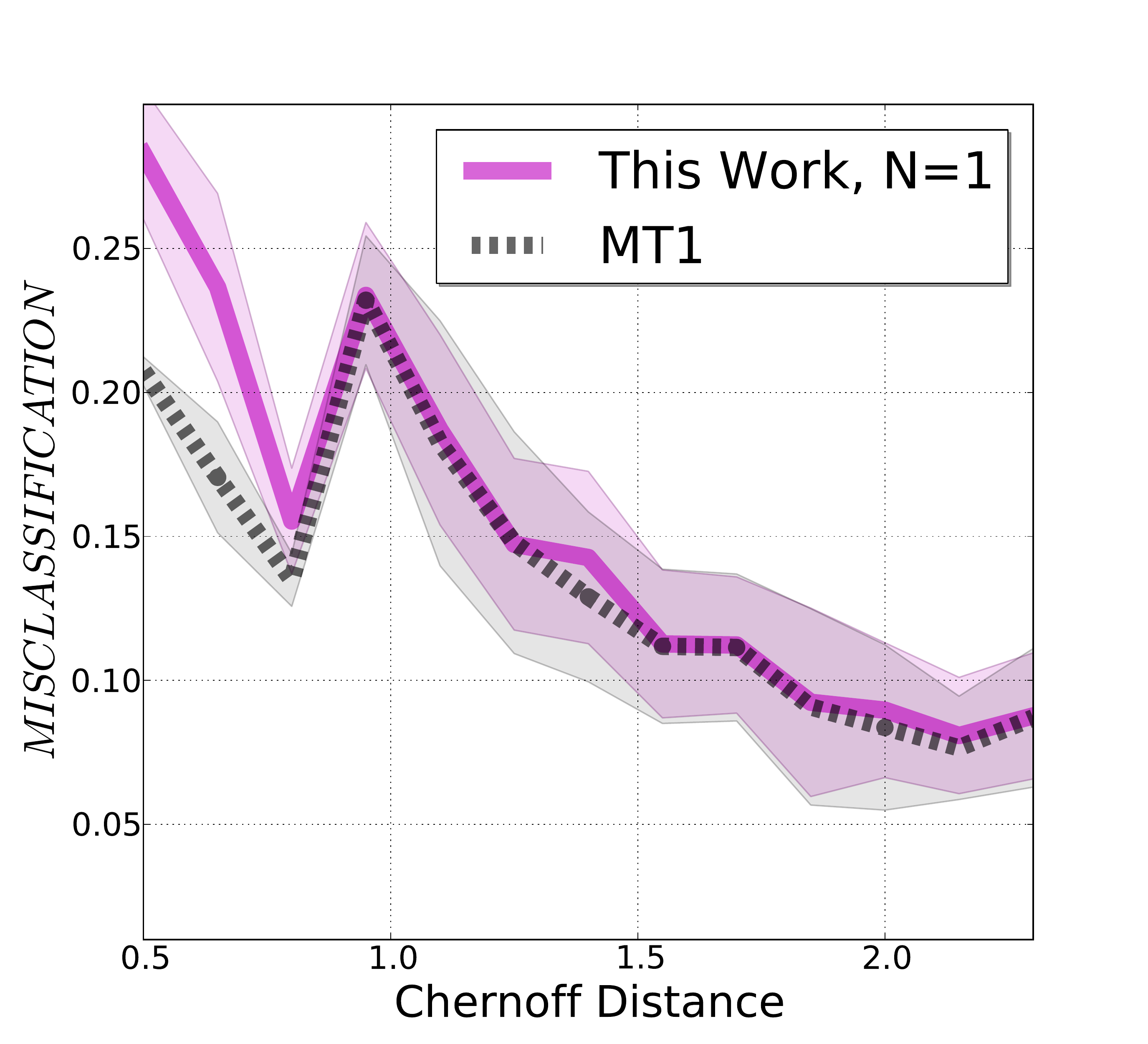} 
\includegraphics[width=7.4cm]{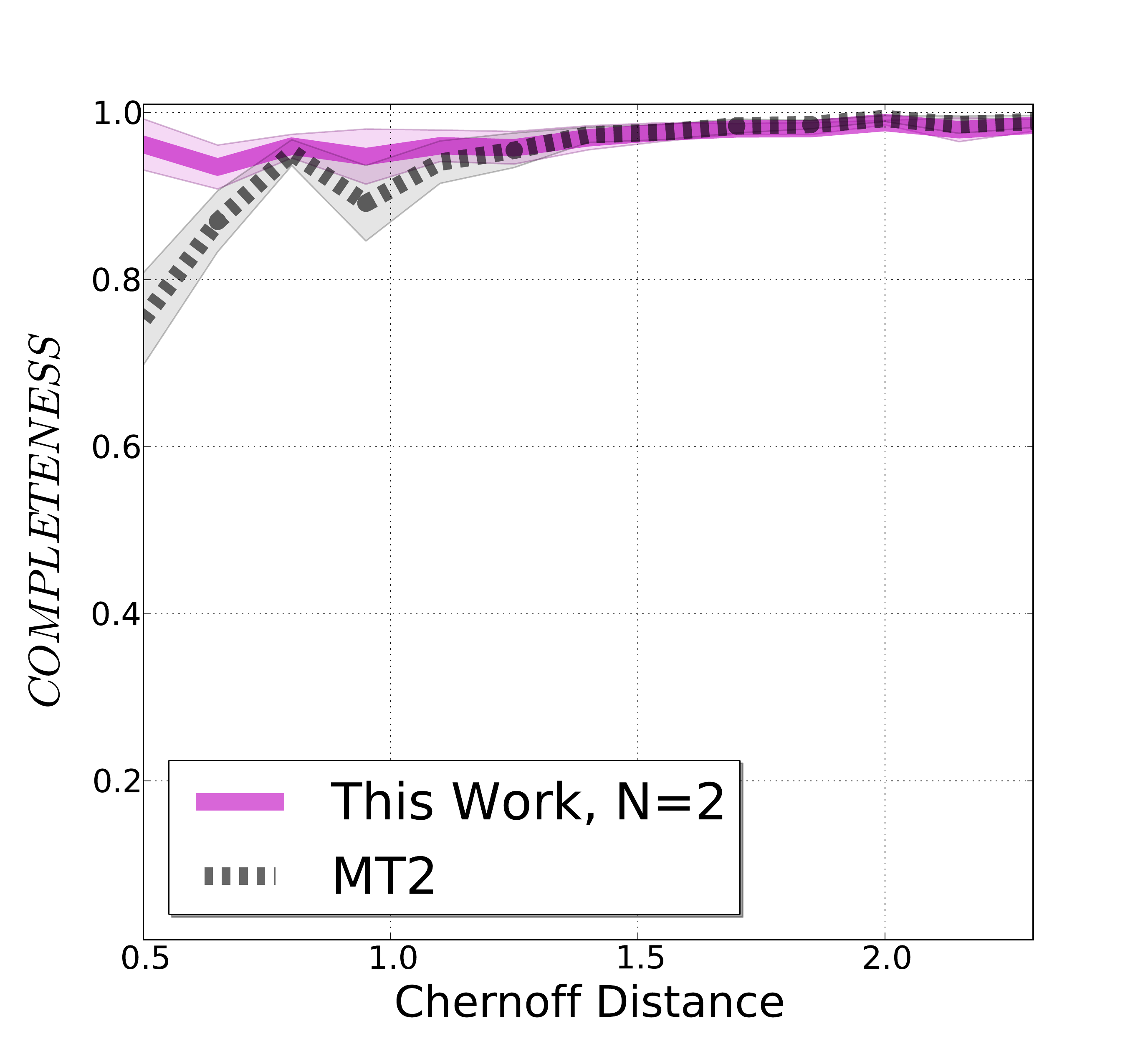} 
\includegraphics[width=7.4cm]{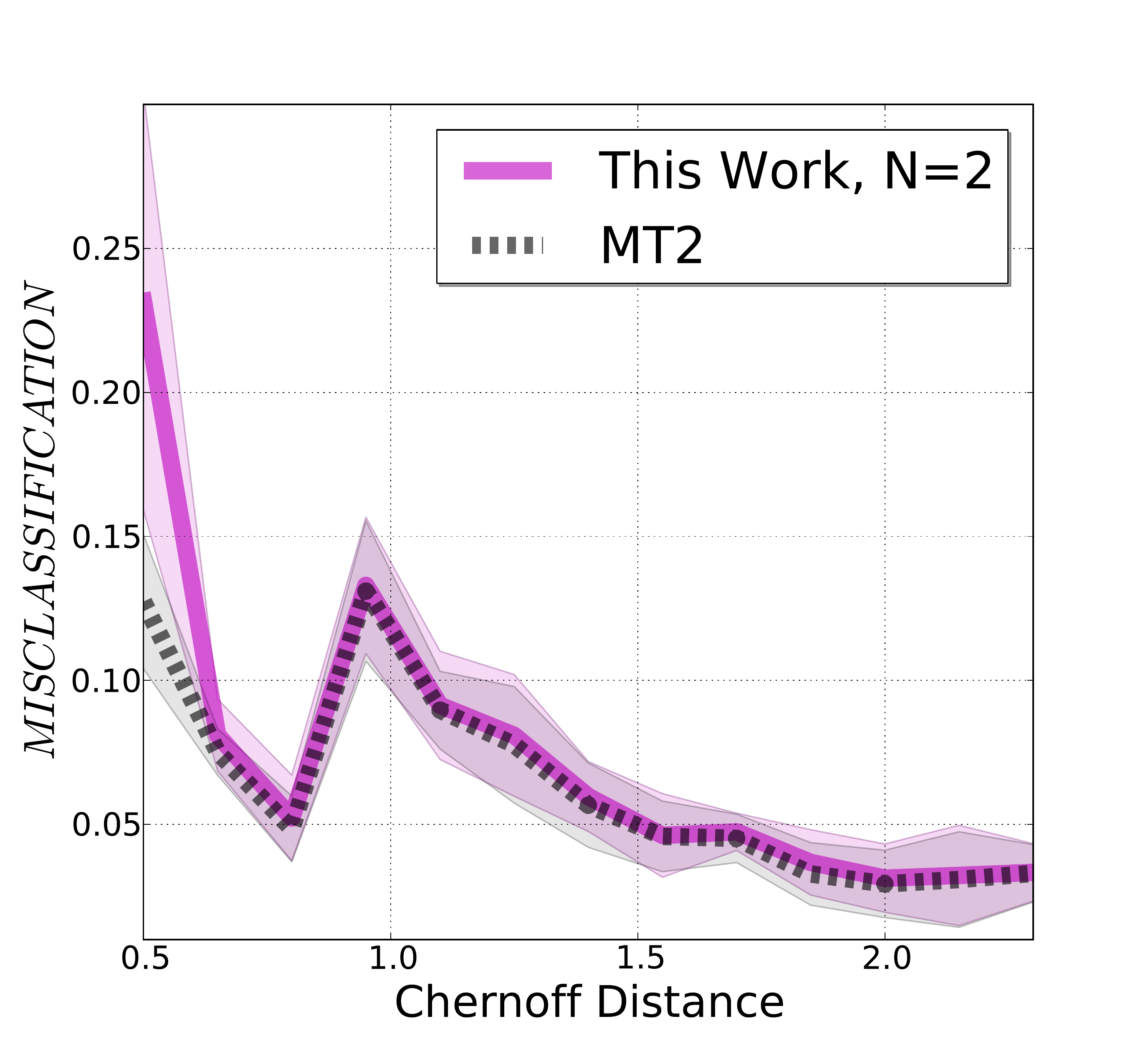} 
\includegraphics[width=7.4cm]{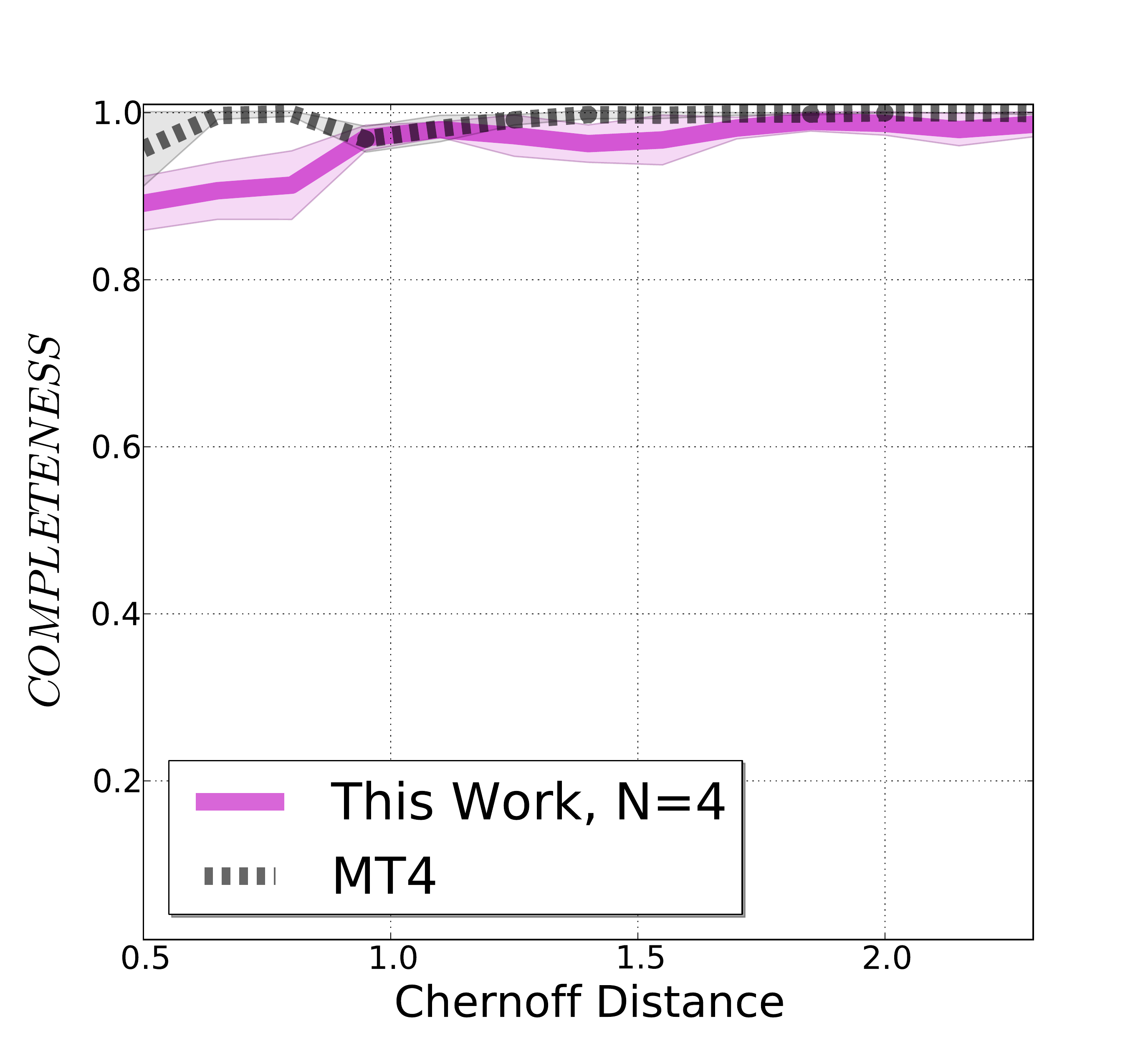} 
\includegraphics[width=7.4cm]{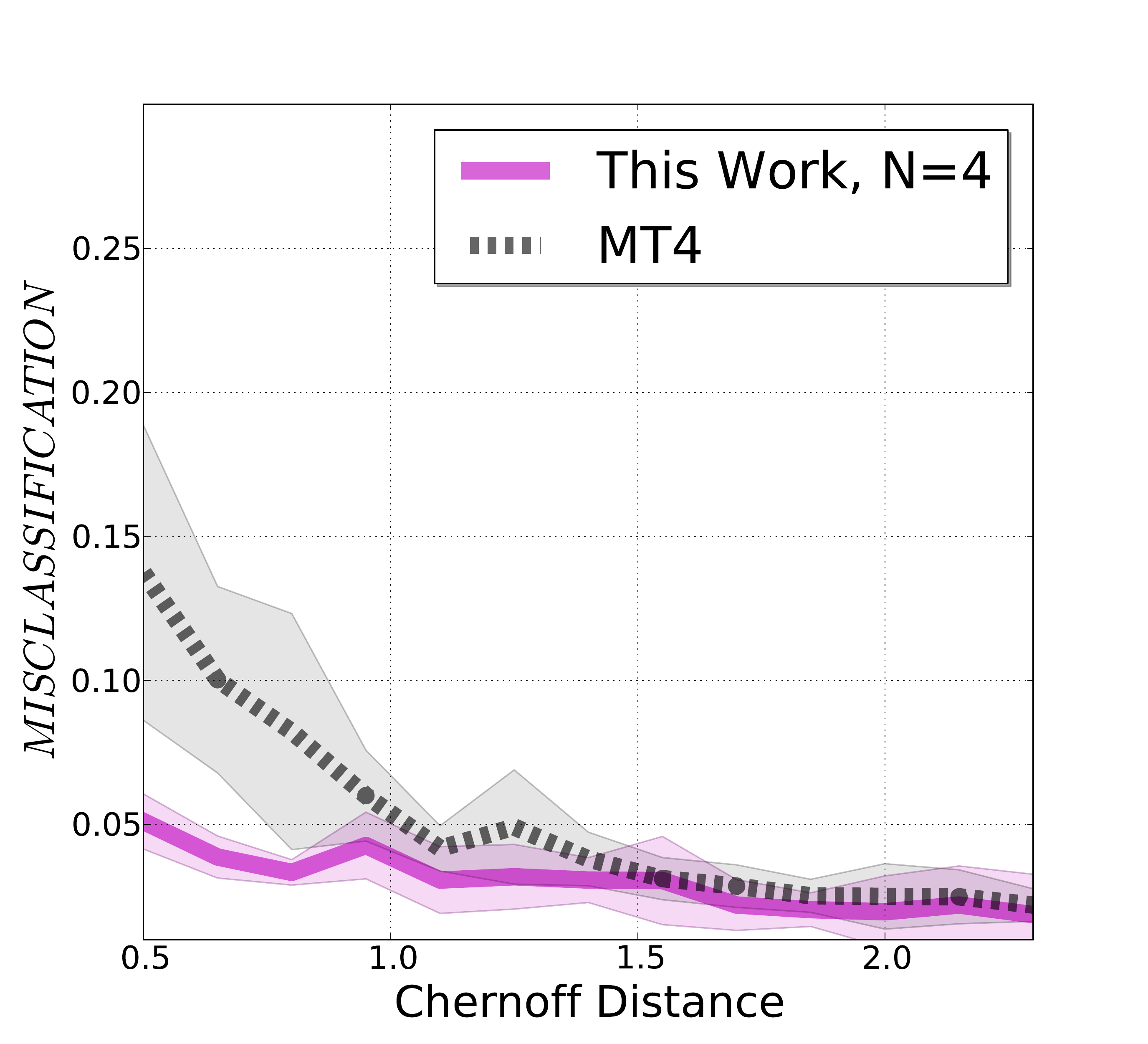} 
\caption{Behaviour of both figures of merit with the number of variables utilised in the membership analysis, according to CD. The continuous magenta lines refer to the results obtained by the new methodology, while the dashed black lines show the results obtained by the other methodologies. The shaded areas represent the dispersion (of 1$\sigma$) of the results within an interval of 0.15 in units of distance. An improvement in the results is observed, both with the increase in the number of variables and with the heteroscedasticity of the distribution functions measured by the CD between the two populations.}
\label{results}
\end{center}
\end{figure*}

Once the outliers have been removed, the different methodologies described above have been applied, and the results detailed below have been obtained. Figure \ref{results} summarises the main results of the comparison of the different methodologies applied to the simulated data. The figure is divided into six panels, the two upper panels analysing the case of one variable (N = 1), the middle panels that of two variables (N = 2), and the lower panels the case of four variables (N = 4). For each set of variables, the graphs show the figures of merit \textit{C} and \textit{M} versus the CD. Take note that the estimation of the CD enables one to incorporate all the parameter variations that the simulated distribution functions describe. Below we shall also analyse the behavior of these figures of merit with singular parameters of the distributions.   

The results of the new method application are indicated by a continuous magenta line, while the results of the model with which it is compared in each case are represented by a dashed black line. The shaded areas represent the dispersion (1$\sigma$ wide) of the results within an interval of 0.15 in units of distance.

In Figure 4 it can be observed that the \textit{M} obtained by all the methodologies decreases upon increasing the number of variables utilised in the analysis. We should note that the \textit{M} rate obtained by the new methodology decreases from 25\% for one variable to 5\% for four variables, for the set of CD values lower than 0.9. Likewise \textit{C} increases with the number of variables utilised, but for the new methodology the \textit{C} can be considered constant (\textit{C} $>$90\%) for any set of variables and within the CD range analysed. It should be noted that for the case of N = 4 variables the \textit{M} corresponding to CDs less than 1 improves with the proposed new methodology if we compare it to methodology MT4. The right-hand column in Figure 4 shows an abrupt change in the behavior of \textit{M} for CD values between 0.9 and 1.0. This behavior, as will be seen further on, is due to the change in the proportion of cluster stars when going from 20\% to 40\%.

From a general point of view, as the CD between cluster and field stars increases, all the methodologies are able to recover a greater number of members, introducing a lower contamination of field stars. The greater differences in the values obtained from \textit{C} and \textit{M} between methodologies are observed in the lowest CD values.

\begin{figure*}
\begin{center}
\includegraphics[width=8.4cm]{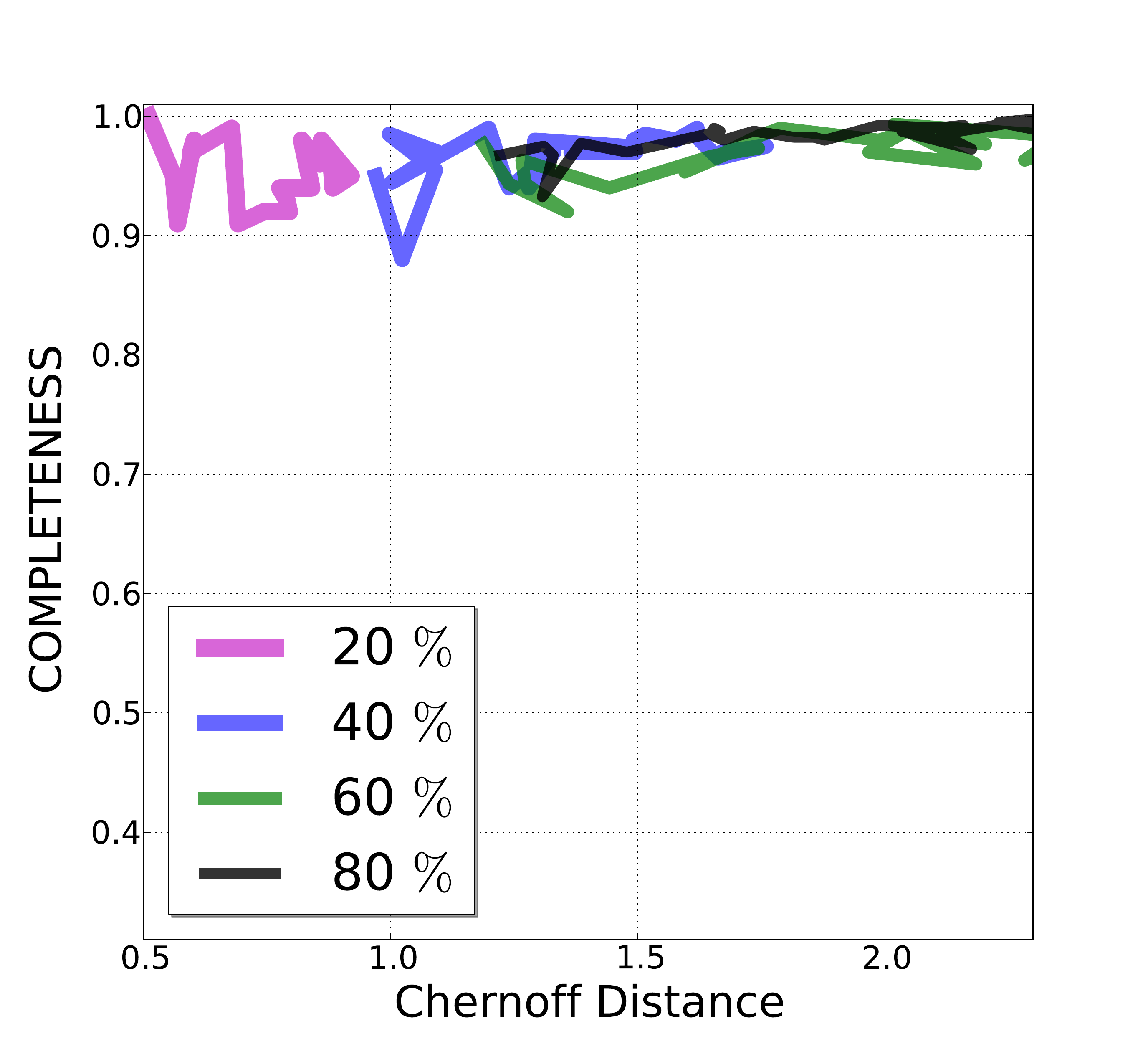} 
\includegraphics[width=8.4cm]{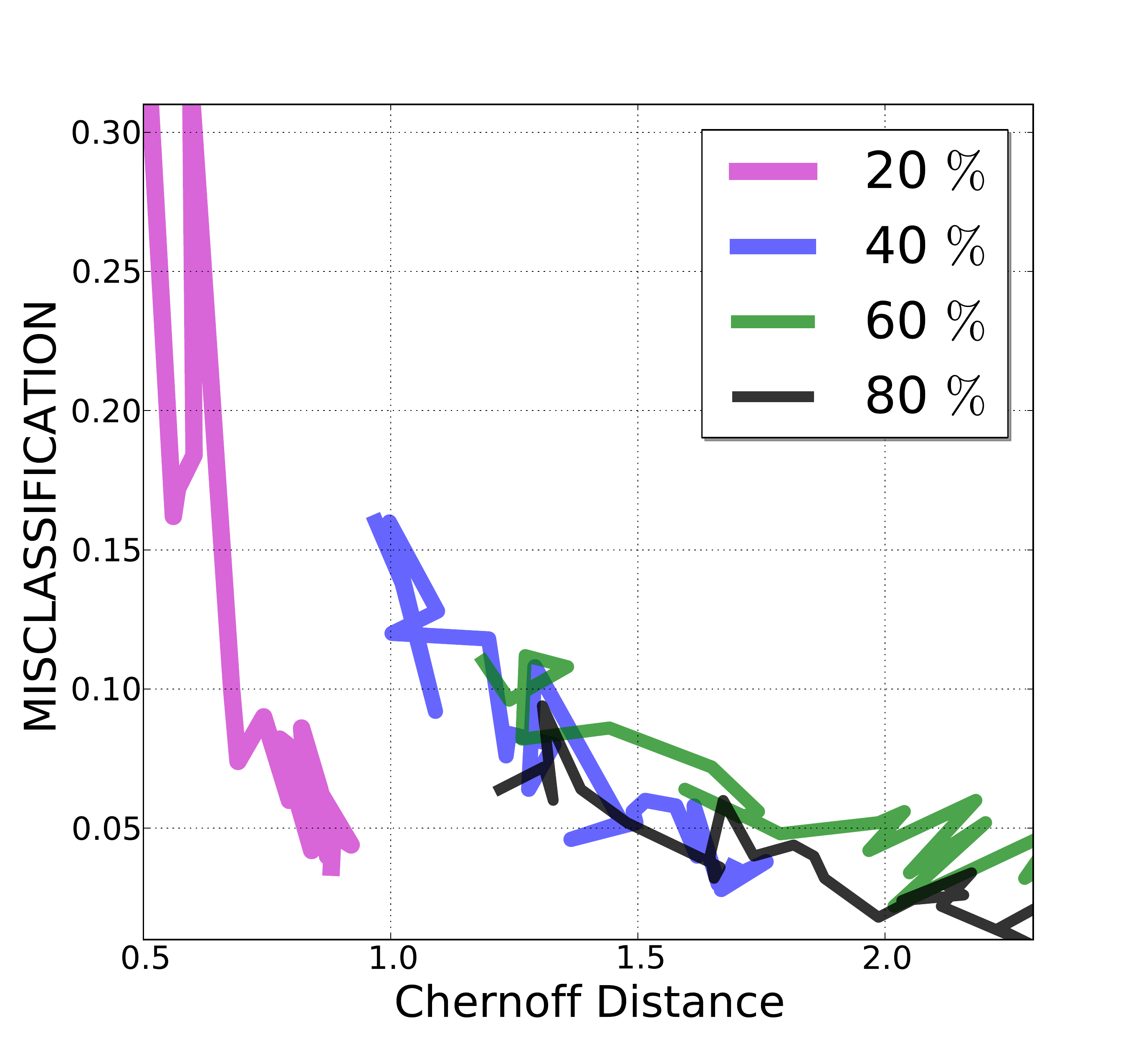} 
\caption{Dependence with the percentage of simulated cluster stars of the values of \textit{C} and \textit{M} according to CD. The results obtained for a number of cluster stars equal to 20\% of the total sample are represented in magenta, for 40\% in blue, 60\% in green and 80\% in black. These results correspond to those obtained by the new methodology for N = 2 variables. The lower CD values (CD $<$ 0.9) correspond with the lowest percentage of cluster stars. For larger percentages of cluster stars the CD increases, with a decrease in the values of \textit{M} also being observed.}
\label{cm_models}
\end{center}
\end{figure*}

The results show that the proportion of the number of cluster stars in the sample also has a large influence on the figures of merit. Figure \ref{cm_models} shows the values of \textit{C} and \textit{M} obtained by the new methodology for N = 2 for the four simulated cluster star percentages, that of 20\% in magenta, 40\% in blue, 60\% in green and 80\% in black. It is observed that the lowest CD values (CD $<$ 0.9) correspond only with the lowest percentage of cluster stars, that of 20\%. Moreover, while \textit{C} has a practically constant behavior, it is observed that the larger the percentage of cluster stars in the sample, the lower the value of \textit{M} obtained.

\begin{figure*}
\begin{center}
\includegraphics[width=8.4cm]{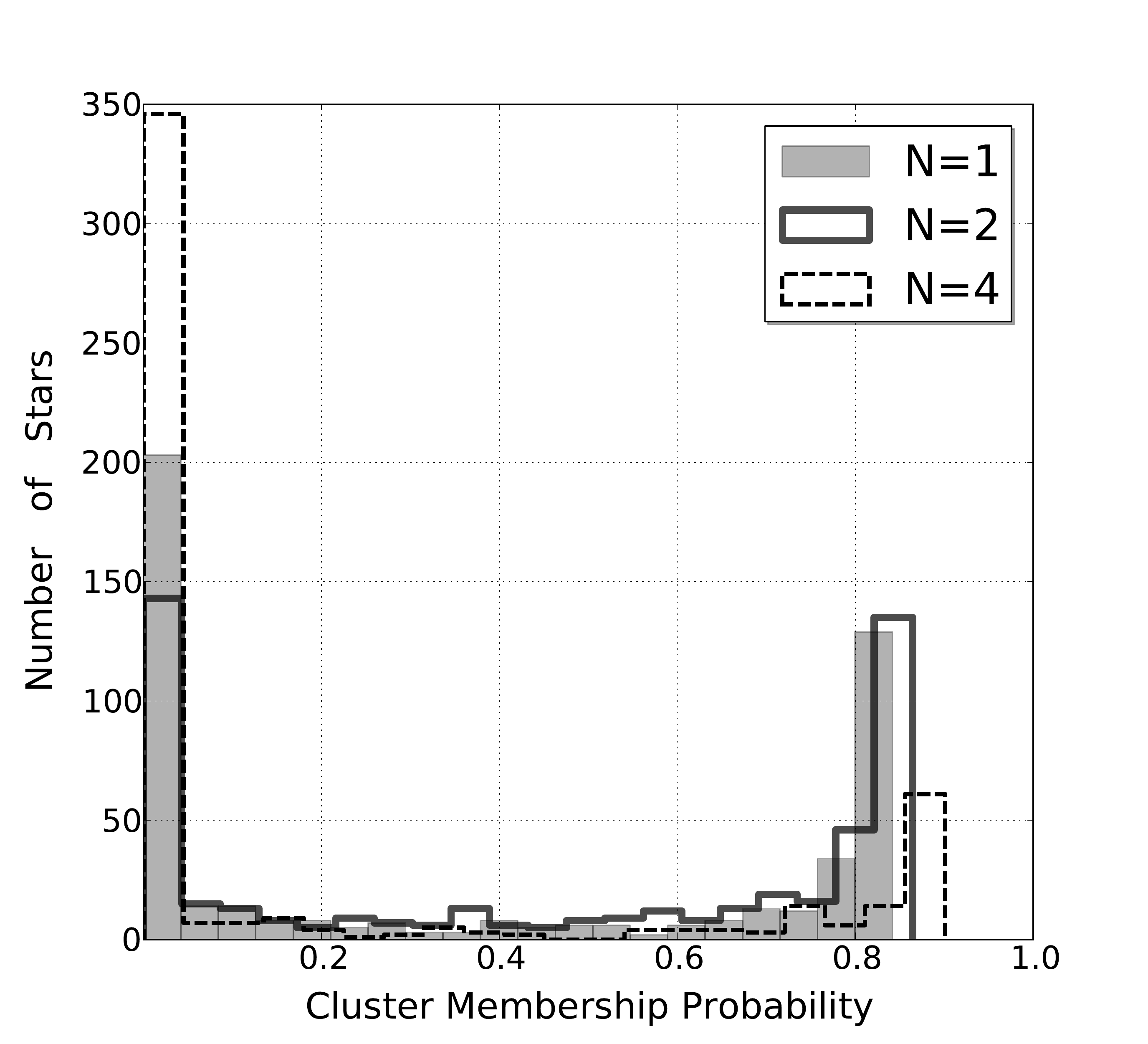} 
\includegraphics[width=8.4cm]{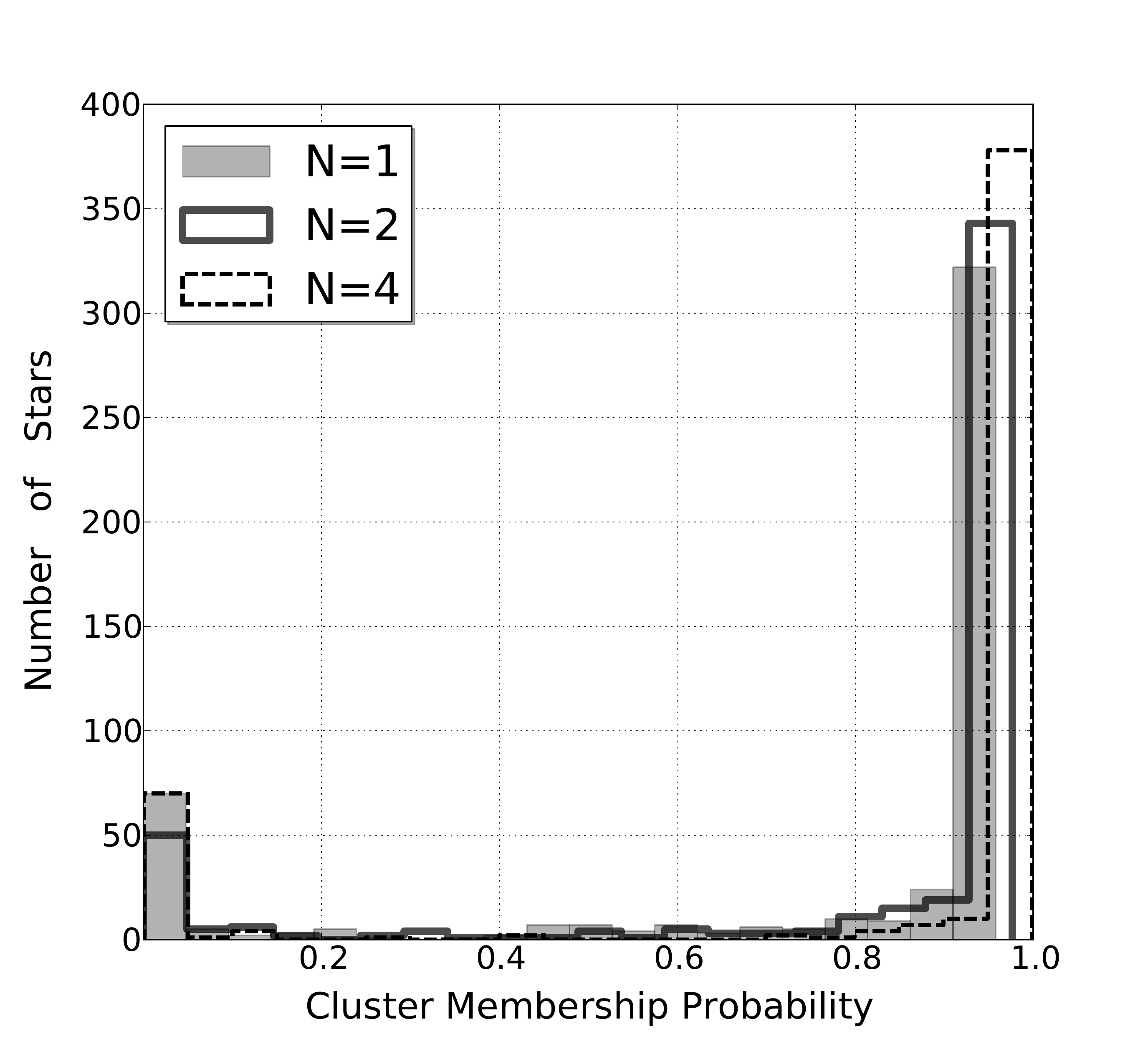} 
\caption{Distribution of the cluster membership probabilities obtained by the methodology proposed in this work, for N = 1, 2 and 4 variables. These results correspond to the same simulated model in which the only variable parameter is the fraction of cluster stars, 20\% on the left, and 80\% on the right.}
\label{histograms}
\end{center}
\end{figure*}

The probability distributions of cluster membership obtained by the new methodology are influenced both by the number of variables utilised in the analysis and by the percentage of cluster stars simulated. In Figure \ref{histograms} the cluster membership probability distributions obtained after applying the new methodology for N = 1, 2 and 4 are shown. These results correspond to a number of cluster stars equal to 20\% of the sample total, on the left, and to 80\% of the sample on the right, maintaining the same values in the rest of the parameters that define the distributions of both populations. It is observed that for a 20\% proportion of stars the highest probability attained is of the order of 90\%. For both cases, the highest values of membership probability are those obtained for N = 4.

To find out the uncertainties in the estimation of the two figures of merit obtained by the new methodology for N = 1, 2 and 4 variables, the Bootstrap technique \citep{Boot} is applied. For this, 50 resamplings of each of the simulated models were carried out, obtaining uncertainties of $\sim$ 1\% for both \textit{C} and \textit{M}.

\begin{figure*}
\begin{center}
\includegraphics[width=8.4cm]{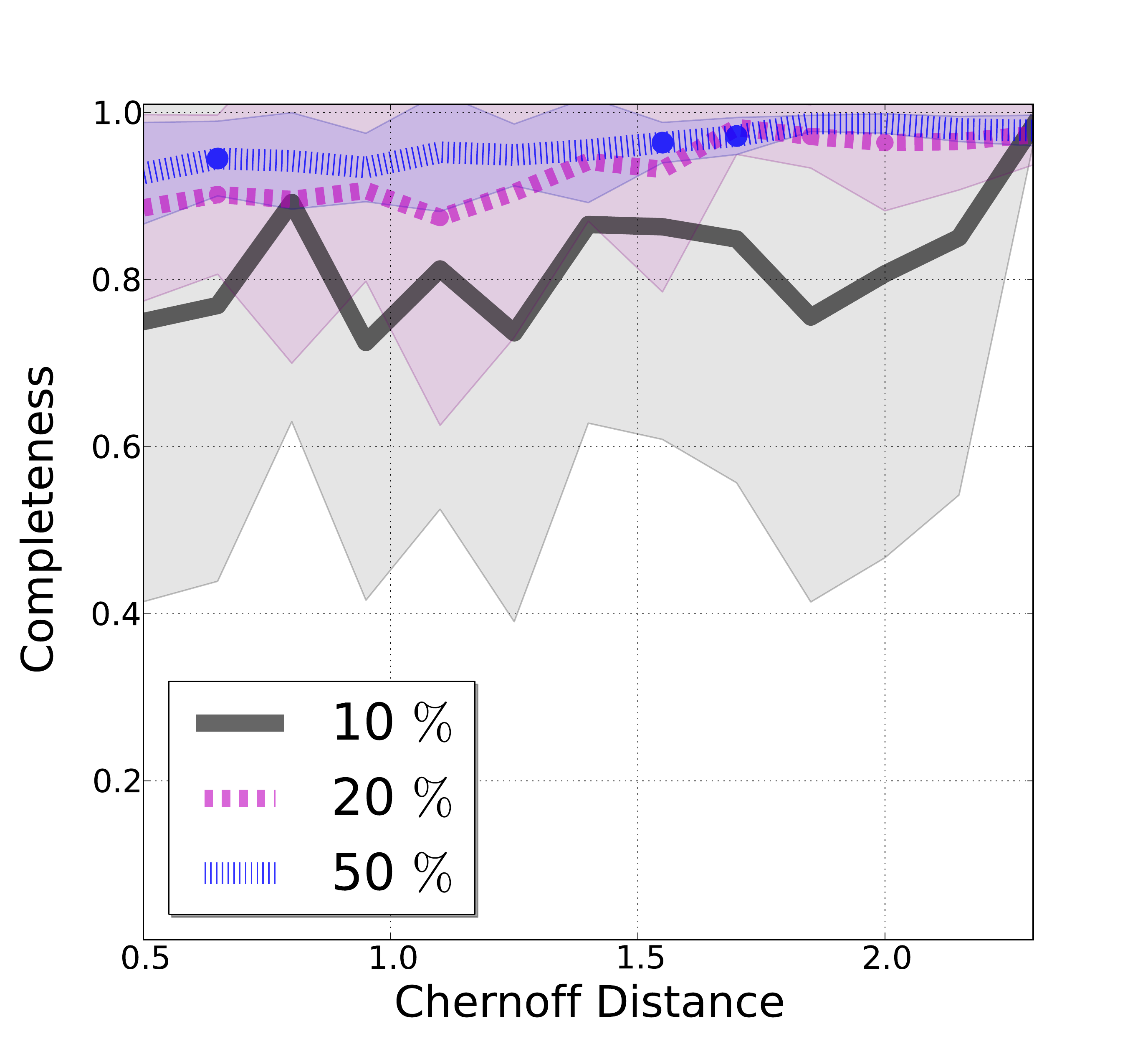} 
\includegraphics[width=8.4cm]{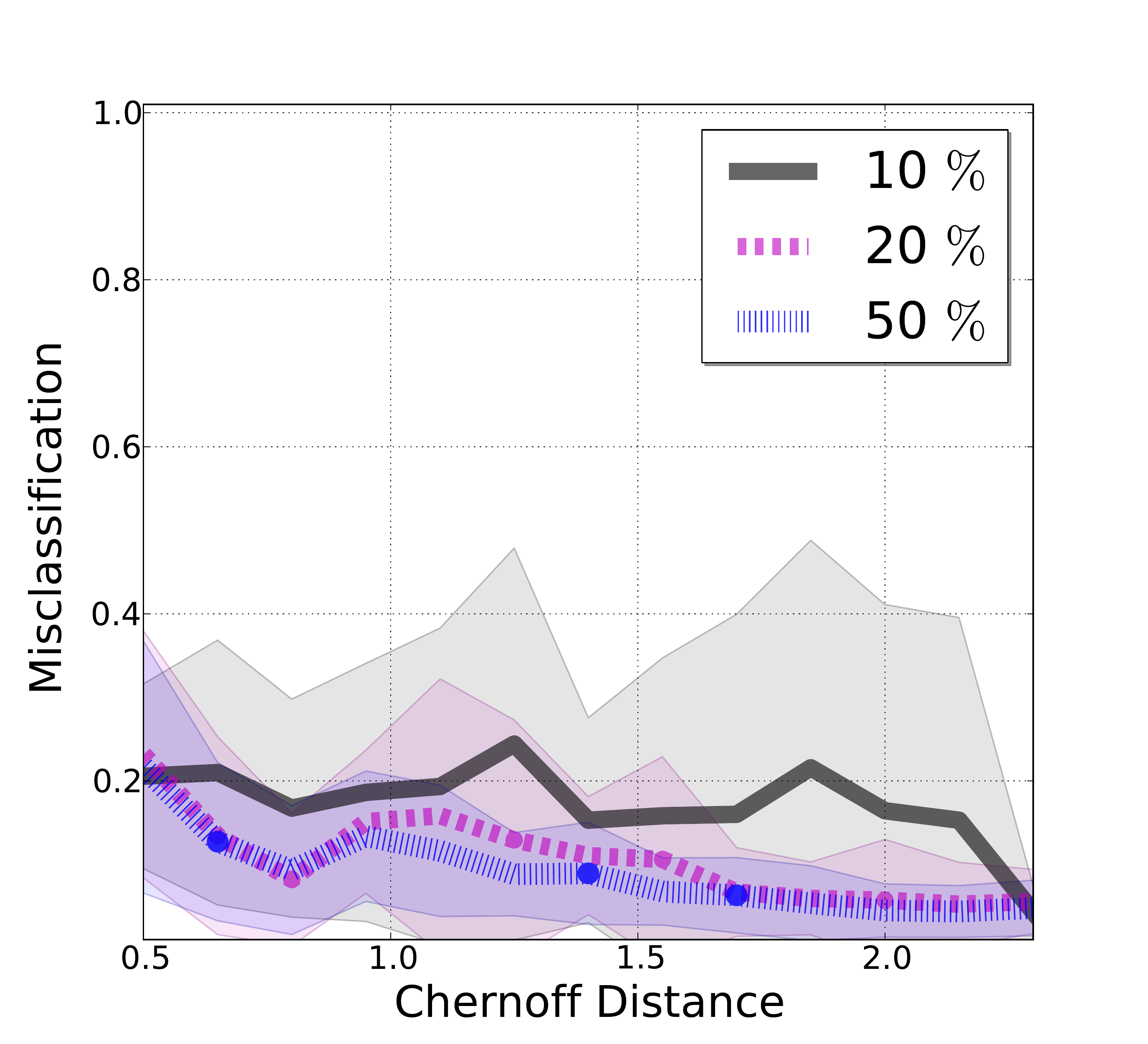} 
\caption{Influence on the determination of \textit{C} and \textit{M} of the different simulated observational subsamples. The results referring to the 50\%, 20\% and 10\% samples are shown by a close-dashed blue line, a wide-dotted magenta line and a continuous black line, respectively.}
\label{cm_baja_estadistica}
\end{center}
\end{figure*}

The results obtained by the new methodology after their application to the different observational samplings of the phase-space distribution functions are presented in Figure \ref{cm_baja_estadistica}. The figure-of-merit values obtained are better, on average, in the samplings of 50\% (close-dashed blue line) and 20\% (wide-dotted magenta line) for any of the proportions of cluster stars and number of variables utilised. Although the average values obtained in the 10\% sampling (continuous black line) are not very different to those obtained by the other two subsamples, the uncertainty in the estimation of C and M is greater, which affects the correct classification of the sample into cluster and field. This can be observed from the coloured zone (1 $\sigma$) associated with this subsample. This result shows the importance of having a good completeness in the catalogues utilised for the correct classification of the sample into the two stellar populations.

Tables \ref{table_pos} and \ref{table_pm} show the differences between the means and the dispersions obtained after the application of the new methodology using N = 1, 2 and 4 variables, with respect to the simulated models. The results of the three subsampling cases are shown in detail, as well as the initial test cases in which there was a total of 500 stars. It can be seen that the use of a larger number of variables in the membership analysis results in a better attainment of the cluster parameters. Moreover, if we compare the results obtained for the different subsamples with the initial one, we observe a gradual worsening in the determination of the simulated models' cluster parameters.

\begin{table*}

\caption{Means and dispersions of the differences between the simulated spatial parameters and those obtained by the new methodology for N = 1, 2 and 4 variables, for the four subsampling cases.}
\label{table_pos}
\scalebox{1.1}[1.1]{
\begin{tabular}{|l!{\vrule width 1.2pt}l|c|c|c|c|c|c|}

     \hline
        \textbf{Sampling}                         &	               \textbf{N¼ Var}                     &	                                                                        \multicolumn{4}{|c|}{Cluster Spatial Coordinates (deg)}     \\ 
     \hline
     \hline
                                                                         
	                                                   &                  &   \textbf{$\Delta\overline{x_c}$}, \textbf{$\sigma_{\Delta\overline{x_c}}$}  & \textbf{$\Delta{\overline{\sigma_{x_c}}}$}, \textbf{$\sigma_{\Delta\overline{\sigma_{x_c}}}$} &    \textbf{$\Delta\overline{y_c}$}, \textbf{$\sigma_{\Delta\overline{y_c}}$} &   \textbf{$\Delta\overline{\sigma_{y_c}}$}, \textbf{$\sigma_{\Delta\overline{\sigma_{y_c}}}$}      \\         
     \hline
        \textbf{100\%}                     &                              &                                                         &                                      &                                     &        \\
	
 	\hline
	&\textbf{1 Var}                                                  &    0.000, 0.003                &     -0.019, 0.011     &      0.000, 0.003     &    -0.019, 0.011    \\
	\hline
	&\textbf{2 Var}                                                  &    0.000, 0.002                 &    -0.011, 0.009     &       0.000, 0.002     &   -0.011, 0.009   \\
	\hline 
	&\textbf{4 Var}                                                  &    0.000, 0.001               &     0.001, 0.001      &       0.000, 0.001      &   0.001, 0.001    \\

	\hline
       	 \textbf{50\%}                    &                            &                                                       &                                      &                                     &                                  \\
	\hline	      

	&\textbf{1 Var}                                                  &    0.000, 0.004                &    -0.019, 0.012      &     0.000, 0.004     &   -0.019, 0.012     \\
	\hline
	&\textbf{2 Var}                                                  &    0.001, 0.004                &    -0.011, 0.009      &     0.000, 0.003     &   -0.011, 0.009    \\
 	\hline
	&\textbf{4 Var}                                                  &    0.001, 0.003                &     0.001, 0.002      &     0.001, 0.002     &    0.001, 0.002     \\

	\hline
       	\textbf{20\%}                  &                           &                                                          &                                      &                                     &                              \\
	\hline	      

	&\textbf{1 Var}                                                  &    0.001, 0.011                &    -0.019, 0.013      &     0.000, 0.009     &   -0.019, 0.013     \\
	\hline
	&\textbf{2 Var}                                                  &    0.001, 0.006                &    -0.010, 0.011      &     0.000, 0.005     &   -0.012, 0.012    \\
	\hline
	& \textbf{4 Var}                                                 &    0.000, 0.006                &     0.002, 0.004      &     0.000, 0.005     &    0.001, 0.004    \\ 

	\hline
       	 \textbf{10\%}                   &                           &                                                          &                                      &                                     &                                  \\
	\hline	      

	&\textbf{1 Var}                                                  &    0.005, 0.028                &    -0.018, 0.019      &     0.000, 0.022     &   -0.021, 0.017    \\ 
	\hline
	&\textbf{2 Var}                                                  &   -0.001, 0.013                &    -0.014, 0.016      &     0.003, 0.015     &   -0.015, 0.017    \\ 
	\hline 
	&\textbf{4 Var}                                                  &    0.002, 0.014                &     0.003, 0.007      &     0.000, 0.010     &    0.002, 0.008    \\ 
     \hline
     \hline

\end{tabular}}

\end{table*}

\begin{table*}

\caption{Means and dispersions of the differences between the simulated kinematic parameters and those obtained by the new methodology for N = 1, 2 and 4 variables, for the four subsampling cases.}
\label{table_pm}
\scalebox{1.1}[1.1]{
\begin{tabular}{|l!{\vrule width 1.2pt}l|c|c|c|c|c|c|}

     \hline
        \textbf{Sampling}             &	  \textbf{N¼ Var}                     &	                                                                        \multicolumn{4}{|c|}{Cluster Kinematic Coordinates (mas/yr)}     \\ 
     \hline
     \hline
                                                                         
	                            &       &   \textbf{$\Delta\overline{\mu_{c,x}}$}, \textbf{$\sigma_{\Delta\overline{\mu_{c,x}}}$} & \textbf{$\Delta\overline{\sigma_{\mu_{c,x}}}$}, \textbf{$\sigma_{\Delta\overline{\sigma_{\mu_{c,x}}}}$} & \textbf{$\Delta\overline{\mu_{c,y}}$}, \textbf{$\sigma_{\Delta\overline{\mu_{c,y}}}$} & \textbf{$\Delta\overline{\sigma_{\mu_{c,y}}}$}, \textbf{$\sigma_{\Delta\overline{\sigma_{\mu_{c,y}}}}$}  \\         
     \hline
        \textbf{100\%}                     &                           &                                                         &                                      &                                     &        \\
	
 	\hline
	&\textbf{1 Var}                                                  &    -0.013, 0.695                &     -5.693, 3.442      &     0.142, 0.316     &    0.031, 0.379    \\
	\hline
	&\textbf{2 Var}                                                  &    0.001, 0.147                &      0.033, 0.272       &     0.093, 0.373     &    -0.114, 0.400    \\
	\hline
	&\textbf{4 Var}                                                  &    -0.025, 0.109                &    -0.147, 0.279       &     0.027, 0.151     &    -0.231, 0.317   \\

	\hline
       	 \textbf{50\%}                    &                               &                                         &                                  &                              &                            \\
	\hline	      

	&\textbf{1 Var}                                                  &   -0.143, 0.846                &    -5.531, 3.634      &     0.244, 0.561     &   -0.039, 0.712     \\
	\hline
	&\textbf{2 Var}                                                  &   -0.020, 0.385                &     0.006, 0.420      &     0.142, 0.433     &   -0.154, 0.534    \\
 	\hline
	&\textbf{4 Var}                                                  &   -0.026, 0.229                &    -0.137, 0.382      &     0.049, 0.289     &   -0.232, 0.391    \\

	\hline
       	\textbf{20\%}                  &                                  &                                         &                                &                             &                              \\
	\hline	      

	&\textbf{1 Var}                                                  &   -0.023, 1.541                &   -5.771, 4.226      &     0.782, 3.086     &   -0.170, 3.454    \\
	\hline
	&\textbf{2 Var}                                                  &    0.012, 0.491                &     0.003, 0.651      &     0.275, 0.759     &   -0.083, 0.928    \\
	\hline
	&\textbf{4 Var}                                                 &   -0.062, 0.517                &    -0.104, 0.496      &     0.137, 0.436     &   -0.119, 0.534    \\ 

	\hline
       	 \textbf{10\%}                   &                                 &                                         &                                &                             &                              \\
	\hline	      

	&\textbf{1 Var}                                                  &    0.763, 5.752                &    -6.262, 6.056      &     2.491, 9.565     &   -0.696, 5.546     \\ 
	\hline
	&\textbf{2 Var}                                                  &    0.722, 3.964                &    -0.651, 2.400      &     1.595, 3.641     &   -0.851, 2.663    \\ 
	\hline 
	&\textbf{4 Var}                                                  &    0.017, 1.839                &    -0135, 1.457      &     0.198, 1.079     &    -0.147, 1.317    \\ 
     \hline
     \hline

\end{tabular}}

\end{table*}

\subsection{Analysis of real data.}

In this subsection, we analyse the application of the new method to real data, specifically to the open cluster NGC 2682. The data used for the analysis comes from the work of \cite{1993A&AS..100..243Z}. Its catalogue contains position and proper motion data for 1046 stars in the region of the cluster. 

To demonstrate the potential of the methodology proposed in this work in the use of different datasets, we have carried out two membership analyses with N = 2 and 4 variables, and compared the results obtained with the two methodologies described above. Thus we have applied methodology MT2 to the proper motions variables, to compare their results with those obtained by the new methodology. This cluster was analysed using the methodology MT4 in the work by \cite{2009ApJ...696.2086S}. We will use the member determination performed in this work to compare with our results. Moreover, the catalogue \cite{1993A&AS..100..243Z} provides us with membership probabilities calculated from the proper motions, which we have used to carry out another comparative analysis, for which we consider a threshold value on the probability of 0.5 to classify the sample into cluster and field stars. 

Prior to the application of the methodologies, the outliers in the proper motion subspace have been estimated, restricting the study to stars with proper motion data lower than $\pm$ 30 mas/yr. This range of proper motions is greater than ten times the cluster proper motions dispersion obtained by \cite{1993A&AS..100..243Z}, which means that a star situated outside this square might be considered as an outlier. This previous selection enables us, on the one hand, to reduce computational time without detriment to the quality and robustness of the results obtained; and, on the other hand, it also facilitates the convergence of the OUTKER procedure. The OUTKER algorithm determines 91 outliers, leaving a sample of 922 stars, which are those used to carry out the membership analysis.   

\begin{figure*}
\begin{center}
\includegraphics[width=8.4cm]{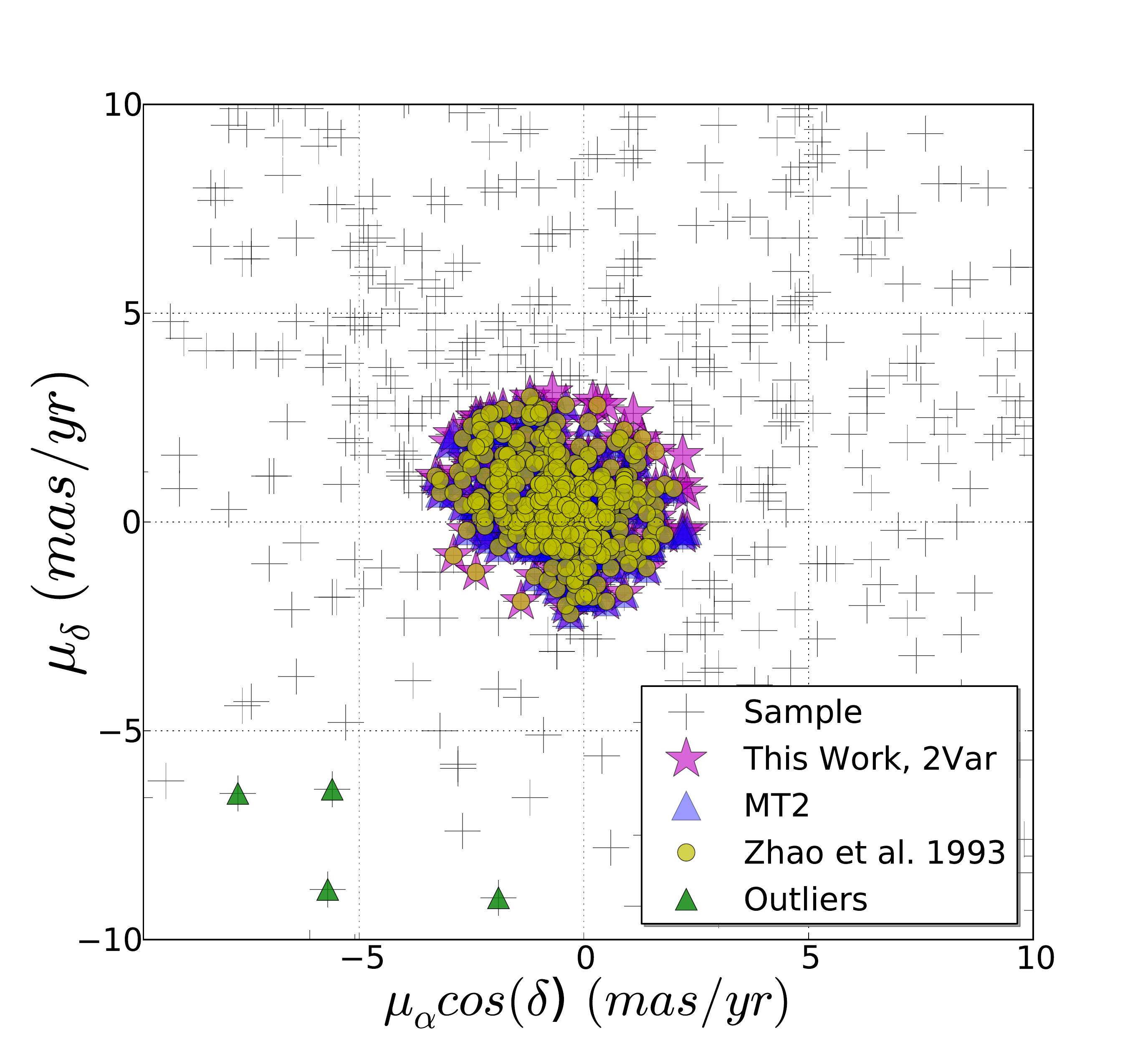}
\includegraphics[width=8.4cm]{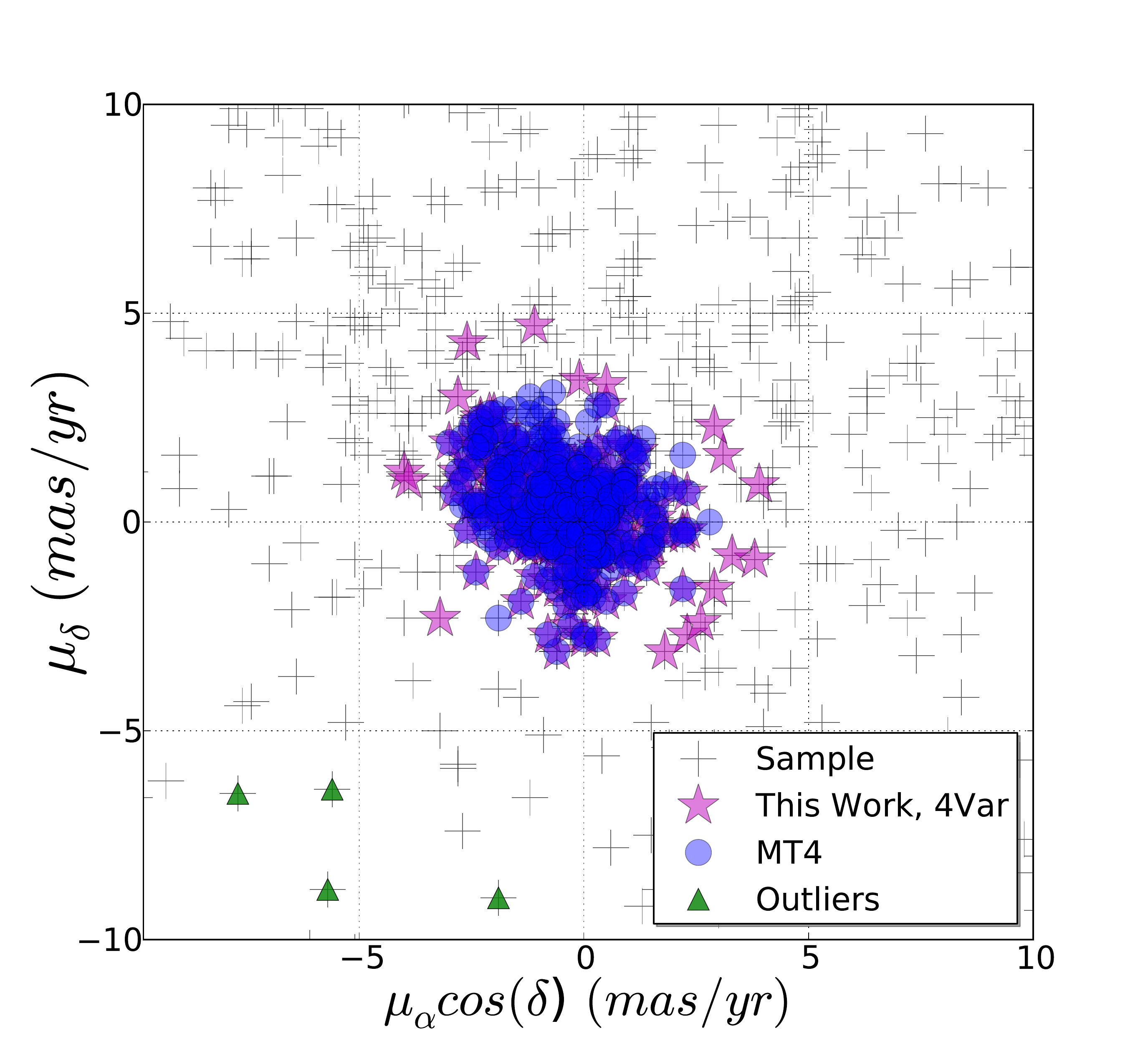}
\caption{Members of cluster NGC 2682 determined by the different methodologies in the proper motion subspace. The graphs have been divided according to the variables utilised in the membership analysis: N = 2 on the left and N = 4 on the right. The sample total is represented in black, the outliers in green, the results obtained by the new methodology in magenta, those obtained by Zhao et al. (1993) in yellow, and those obtained by MT2 and MT4 in blue.}
\label{ngc2682_members}
\end{center}
\end{figure*}

Figure \ref{ngc2682_members} shows the members determined by each of the methodologies in the proper motions subspace. In the graph on the left the members obtained by the methodologies that use the N = 2 variables, and on the right, those using N = 4 variables.

The parameters of the distribution functions that describe the field and cluster populations in the sub-phase space, obtained by each membership analysis, are presented in Table 5. The results show that the number of cluster members obtained by the different membership analyses is very similar, around 356, with the exception of those obtained by the new methodology for N = 4, which determines 314. 

In the positions space, the dispersions of the cluster distribution functions obtained both by the new methodology for N = 4 and by MT4 are lower than those obtained when only the proper motions are used in the membership analysis. Hence the introduction of the positions restricts the cluster determination to more central regions of the positions space. The rest of the parameters that describe the distribution functions of both populations are very similar. 

With regard to the proper motions variables, the cluster parameters determined by the methodologies present a high degree of agreement, with a slight increase in the dispersion obtained in the studies that use N = 4 variables. For the field population, the dispersions obtained by the new methodology and by MT2 are lower than those obtained by \cite{1993A&AS..100..243Z} and MT4. This behaviour is due to the sample utilised to calculate them: both for the methodology MT2 and for the new proposal in this work for N = 2 and 4 variables, the parameters of the field star distribution have been calculated with respect to the sample free of outliers, whereas for both \cite{1993A&AS..100..243Z} and MT4 the calculations have been performed for the total number of stars present in the catalogue.     

The comparison of the number of members common between methodologies is shown in Table 6. As can be observed, the number of common members is very high, although lower for the case of the new methodology for N = 4 variables, as it determines the fewest potential members.  

From the classification of the sample from \cite{1993A&AS..100..243Z} in cluster and field stars, we have obtained a CD of 1.57. Using this value, and making use of the results obtained in our simulations, we have estimated the \textit{C} and \textit{M} expected in the membership analysis carried out for this cluster. From the graphs in Figure 4 for N = 2 and 4 variables, we observe that the values of \textit{C} corresponding to this distance are around 98\% for all the methodologies. The \textit{M} for this distance is around 5\%, that is to say, approximately 52 stars.

\begin{figure*}
\begin{center}
\includegraphics[width=8.cm]{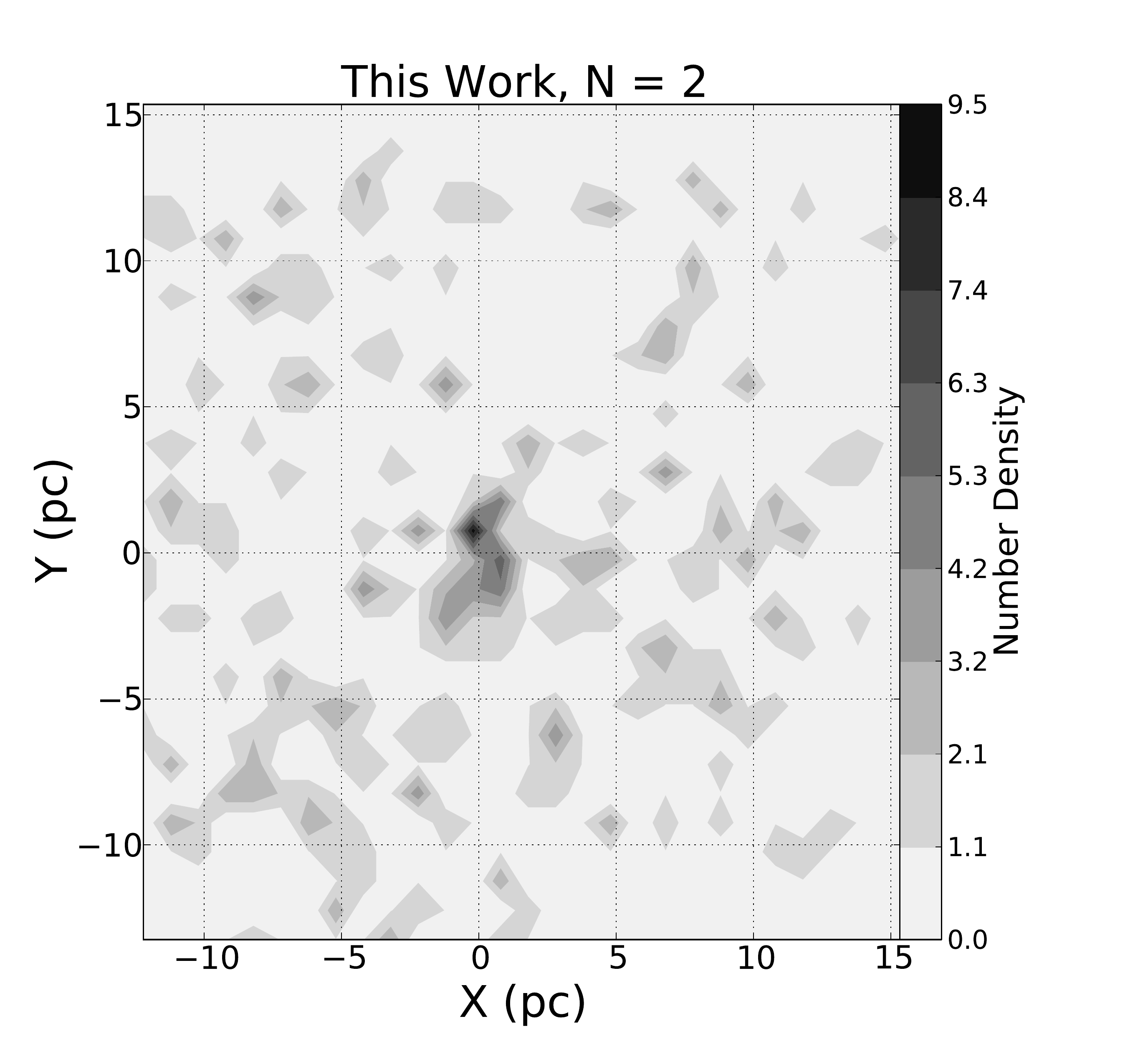}
\includegraphics[width=8.cm]{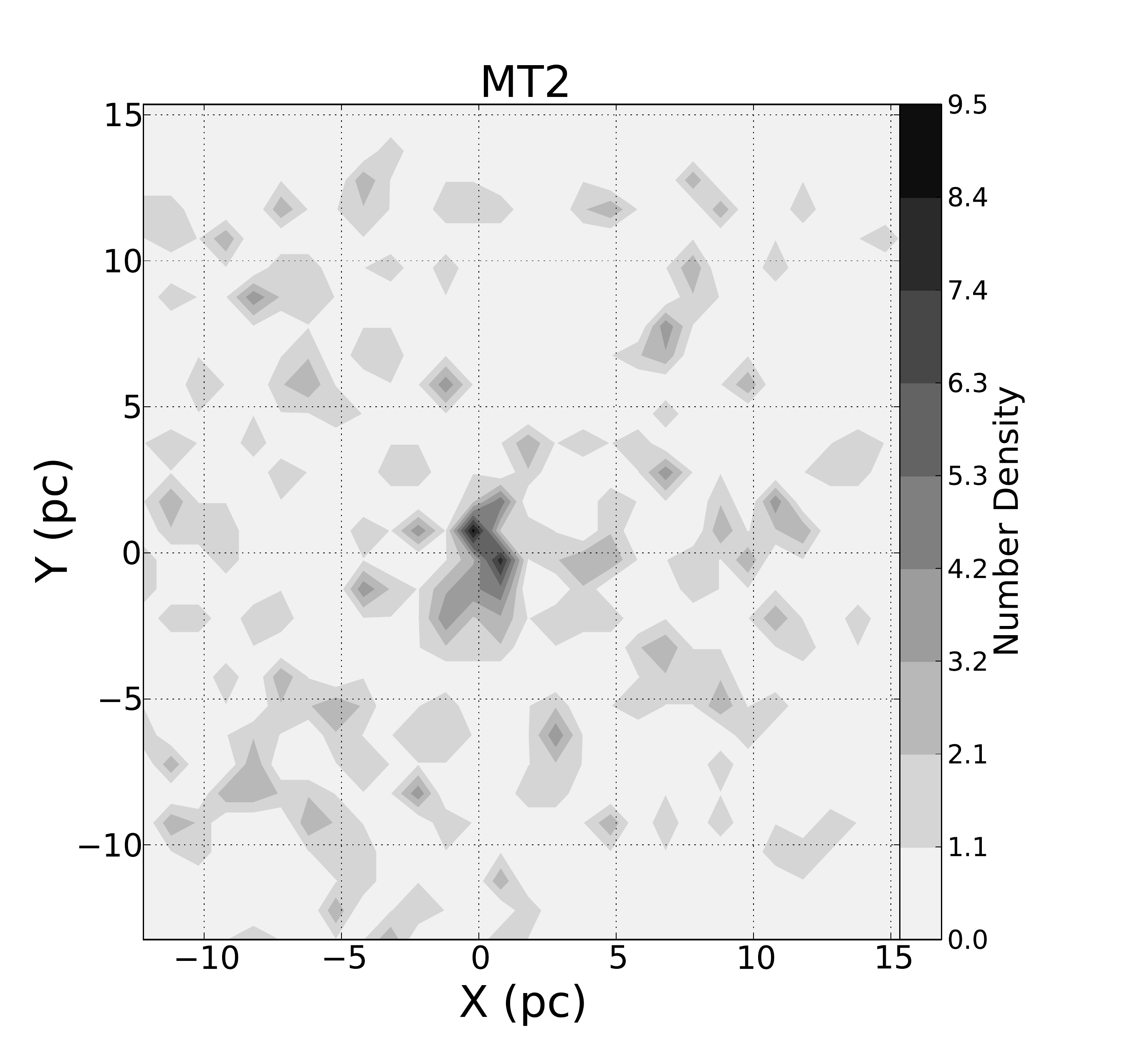}
\includegraphics[width=8.cm]{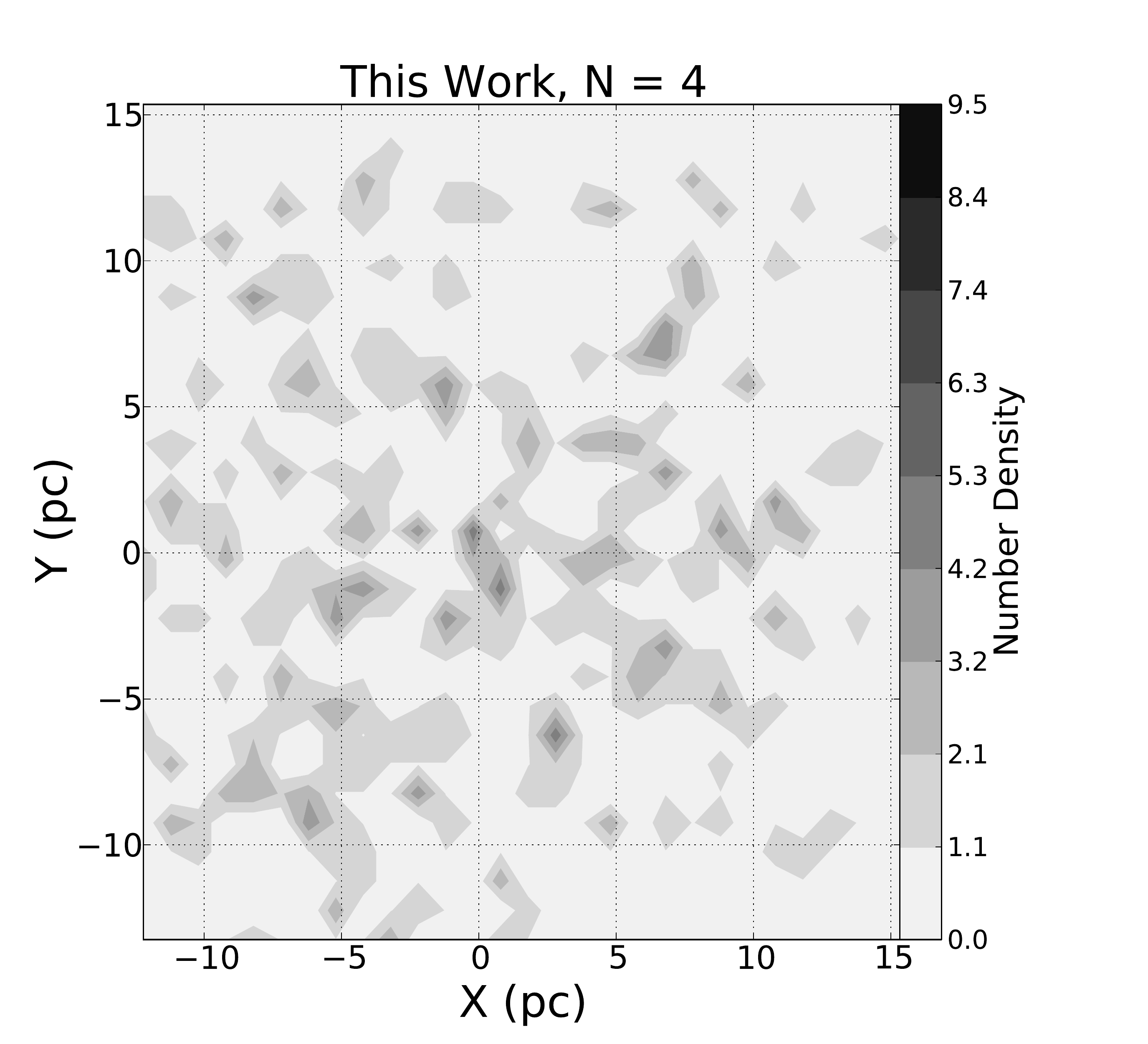}
\includegraphics[width=8.cm]{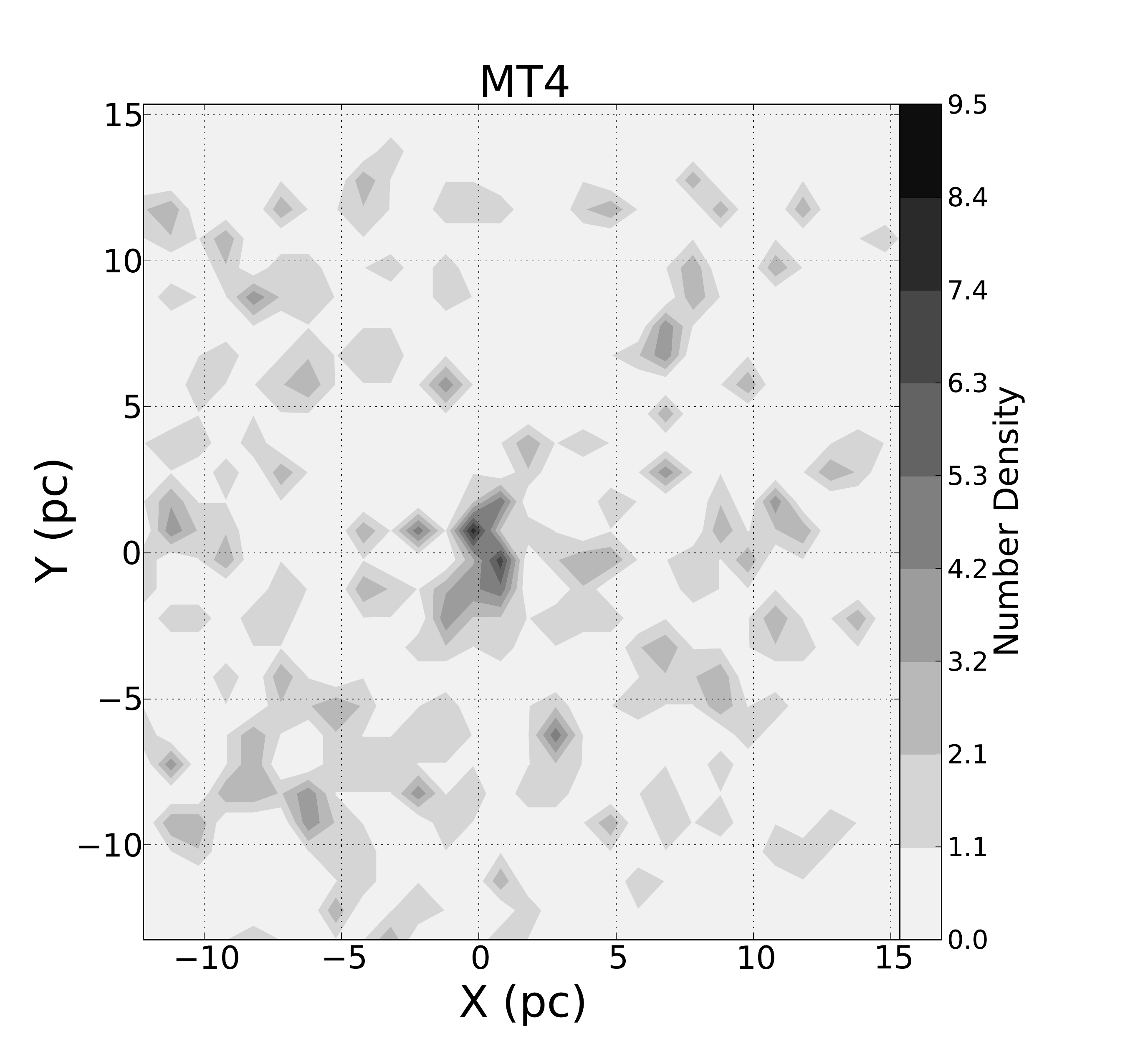}
\includegraphics[width=8.cm]{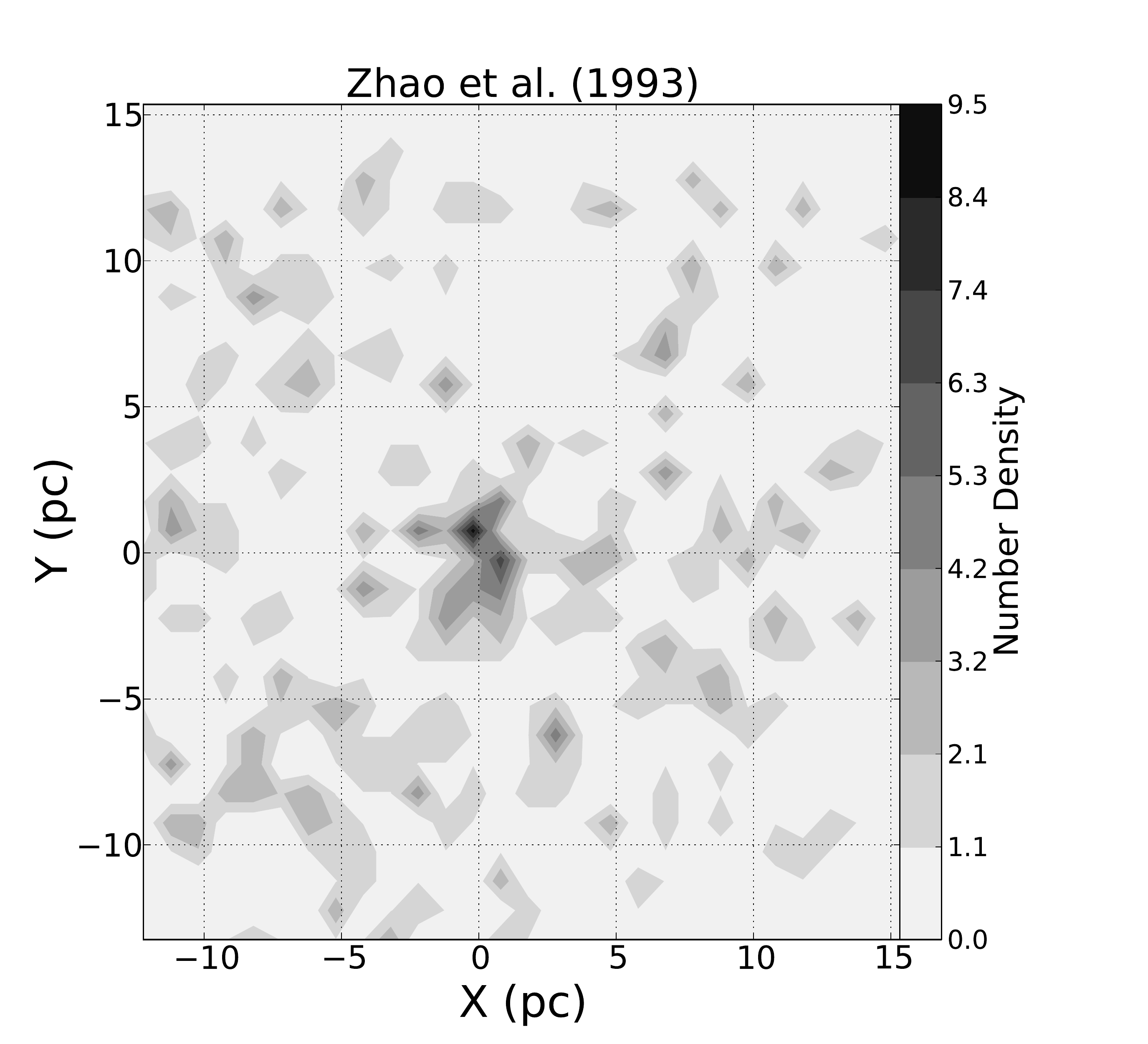}
\caption{Density maps (using a squared beam of 1 pc per side) of the field population determined by different methods. The upper and middle graphs show the distribution of field stars for the membership analyses that utilise the N = 2 and 4 variables, respectively. The lower graph shows the distribution obtained making use of the data from Zhao et al. (1993). Particularly uniform is the distribution of field stars obtained by the new methodology for N = 4 variables, showing the potential of this new methodology in the identification of members in star clusters.}
\label{dens}
\end{center}

\end{figure*}

Given that the cluster can be considered responsible for the overdensity observed in the positions space, after extraction, we should obtain a uniform distribution of field stars. In Figure \ref{dens} the residues obtained are studied after eliminating the cluster determination from the sample performed by each methodology. The density of the field stars, obtained for each methodology, does not show a high degree of structure in the residues, which are perfectly compatible with a discrete homogeneous distribution. Particularly uniform is the distribution of field stars obtained by the new methodology for N = 4 variables, showing the potential of this new methodology in the identification of members in star clusters.

\begin{table*}

\caption{{Parameters of the cluster and field distribution functions obtained by each of the NGC 2682 membership analyses.}}
\label{table_ngc2682}
\scalebox{1.1}[1.1]{
\begin{tabular}{|l!{\vrule width 1.2pt}l|c|c!{\vrule width 1.2pt}c|c|c|c|c|c|}

     \hline
	                                         &	       \multicolumn{5}{|c|}{\textbf{Number of Variables}}      \\         
     \hline
     \hline

	\textbf{Parameters}                                                             &	       \multicolumn{3}{|c|}{\textbf{2 Variables}}          &    \multicolumn{2}{|c|}{\textbf{4 Variables}}   \\         
     \hline
	                                                                                            &\textbf{This Work}&   \textbf{MT2}  &  \textbf{Zhao et al. (1993)} &	\textbf{This Work} &  \textbf{MT4}   \\
	\hline
	Number of Cluster Members                                               &            370           &     356             &             354                      &          314                  &         354   \\ 
 	\hline
	\textit{x$_c$}   (pc)                                                       &            0.224       &      0.055         &              0.052                      &        -0.264             &      0.127     \\
	\hline
	\textit{$\sigma_{x, c}$}   (pc)                                     &            3.078        &     2.893          &            2.821                      &        1.709              &       2.425       \\
	\hline
	\textit{y$_c$}   (pc)                                                       &             1.226       &      1.188          &            1.177                    &          1.295             &       1.265      \\
	\hline
	\textit{$\sigma_{y, c}$}   (pc)                                   &             3.010        &      2.883          &            2.885                      &        1.765             &         2.534       \\
	\hline
	\textit{x$_f$}   (pc)                                                       &            0.266        &      0.371          &           0.243                      &         0.513             &        0.204     \\
	\hline
	\textit{$\sigma_{x, f}$}   (pc)                                     &            7.324        &      7.295          &           7.286                      &        7.264              &       7.359       \\
	\hline
	\textit{y$_f$}   (pc)                                                       &              0.872       &      0.904          &            0.965                    &          0.868             &       0.920      \\
	\hline
	\textit{$\sigma_{y, f}$}   (pc)                                   &               7.271        &      7.231          &           7.413                      &        7.204             &       7.476       \\
 	\hline

	\textit{$\mu_{\alpha,c}$}   (mas yr$^{-1}$)                 &            -0.530       &      -0.566         &           -0.561                      &        -0.426             &      -0.476     \\
	\hline
	\textit{$\sigma_{\mu_{\alpha,c}}$}   (mas yr$^{-1}$)  &             1.141        &      1.104          &            1.076                      &        1.248              &       1.101       \\
	\hline
	\textit{$\mu_{\delta,c}$}   (mas yr$^{-1}$)                  &              0.505       &      0.470          &            0.471                    &          0.266             &       0.380      \\
	\hline
	\textit{$\sigma_{\mu_{\delta,c}}$}   (mas yr$^{-1}$)   &             1.057        &      1.016          &            1.033                      &        1.154             &       1.196       \\
	\hline
	$\rho_c$                                                                                &          -0.246        &     -0.350            &            -0.272                  &           -0.315          &        -0.258    \\ 
         \hline

	\textit{$\mu_{\alpha,f}$}   (mas yr$^{-1}$)                  &              0.568      &      0.563           &           -0.703                    &         0.413           &       -0.747     \\
 	\hline
	\textit{$\sigma_{\mu_{\alpha,f}}$}   (mas yr$^{-1}$)   &              7.413       &      7.326           &            11.507                    &          7.078            &       11.500       \\
	\hline
	\textit{$\mu_{\delta,f}$}   (mas yr$^{-1}$)                   &             4.146        &      4.078           &             2.748                     &         3.935           &       2.794      \\
	\hline
	\textit{$\sigma_{\mu_{\delta,f}}$}   (mas yr$^{-1}$)    &             6.448        &      6.387           &             11.622                     &        6.164           &        11.610     \\
	\hline
	$\rho_f$                                                                                 &          -0.213        &     -0.211            &            -0.150                  &           -0.201          &        -0.149    \\ 
         \hline
     \hline

\end{tabular}}

\end{table*}

\begin{table*}

\caption{{Number of members of NGC 2682 common between methodologies.}}
\label{table_members}
\scalebox{0.92}[1]{
\begin{tabular}{|l!{\vrule width 1.pt}l|c|c|c!{\vrule width 1.pt}c|c|c|c|c|c|}
     \hline
     \hline
     \textbf{Number of Variables}                                               &\textbf{Reference}      &	       \multicolumn{3}{c}{\textbf{2 Variables}}          &    \multicolumn{2}{|c|}{\textbf{4 Variables}}   \\ 
     \hline
	 &                                                                                   &	 \textbf{This Work, N = 2}     &   \textbf{MT2}             &  	 \textbf{Zhao et al. (1993)}     &	 \textbf{This Work, N = 4}   &	\textbf{MT4}    	 \\
     	\hline
     \multirow{3}{*}{\textbf{2 Variables}}
        &\textbf{This Work, N = 2}                                            &      	            ---	                        &  	 	 	   354                           &	                        352                  	        &	                        288          &	               343                 	          \\
	&\textbf{MT2}                                                   &      	           354                       &  	 	 	    ---                            &	                        345                            &	                        284          &	               335                 	          \\
	&\textbf{Zhao et al. (1993)}                                           &      	           352	               &  	 	 	   345                           &	                         ---                  	        &	                        282          &	               333                 	           \\  	           
    	\hline
     \multirow{2}{*}{\textbf{4 Variables}}
        &\textbf{This Work, N = 4}                                            &     	           288                       &  	 	 	   284                           &	                        282                  	         &	                        ---            &	                       295                 	          \\
       & \textbf{MT4}                                           &      	           343                       &  	 	 	   335                           &	                        333                  	         &	                        295          &	               ---                 	          \\

	\hline
	\hline

\end{tabular}} 

\end{table*}

\section{Summary and Conclusions}
\label{intro}

We present a new geometrical method aimed at determining the members of stellar clusters. The methodology computes the distances between every star and the cluster central overdensity, in an N-dimensional space. Through an iterative Wolfe estimation procedure, the membership probabilities for every star in the sample are computed fitting the distance distribution through a mixture of two 1-D Gaussians: one for the cluster members and another for the field stars. After imposing a decision criteria of 0.5 on the probability value, the cluster members are determined. 

The method can handle different sets of variables, which have to satisfy the simple condition of being more densely distributed for the cluster members than for the field stars (as positions, proper motions, radial velocities, abundances or/and parallaxes, are). Thus we designed a series of realistic simulations, in the positions and in the proper motions subspaces populated by clusters and field stars. The simulations not only enable us to quantify how the method is able to distinguish between both populations under different numbers of variables (N), but also to compare the results with those obtained by other existing methodologies, always using the same simulated dataset.

The results obtained have been described according to two figures of merit, \textit{C} and \textit{M}, from which we can quantify how each methodology has classified an initial sample of cluster and field stars. The goodness of the classifications depends on the characteristics of the distribution functions of both populations that is, the heteroscedasticity of the pdfs. A measure of this is the Chernoff Distance (CD), which has been utilised to represent both figures of merit.    

The results show that the increase in the number of variables utilised produces better results, recovering a higher percentage of cluster stars with a lower contamination of field stars. \textit{C} and fundamentally \textit{M} improve with the heteroscedasticity of the pdfs, measured by the CD between the two populations.  

The new methodology produces similar or even better results for \textit{C} than those obtained by other methodologies for any set of variables, obtaining values superior to 90\% for practically all the simulations performed. The \textit{M} obtained by the methodology proposed in this work is greater than that obtained by methodologies MT1 and MT2, but lower than that obtained by methodology MT4 for the shorter CD.  

The proportion of cluster stars in the sample also has a large influence on the figures of merit. It is observed that the lowest CD values correspond only with the lowest percentage of cluster stars, that of 20\%.  The larger the percentage of cluster stars in the sample, the lower the value of \textit{M} obtained by any methodology. For the case of an 80\% proportion of cluster stars \textit{M} is almost constant for any CD. 

We have estimated the error of both figures of merit obtained by the new methodology for variables N = 1, 2 and 4. For this we have applied the Bootstrap method, sampling each model 50 times, and thus obtaining errors of $\sim$1\% in the estimation both of \textit{C} and \textit{M}. 

The results obtained after applying the new methodology to different observational subsampling of the same simulated phase-space distribution function show a gradual worsening in the recovery of the parameters that describe the initial cluster and field populations. In the specific case of the proper motions, the differences obtained between the means are much lower than the mean error introduced in the simulations, which was 3 mas/yr. 

As a practical example, the new methodology has been applied to cluster NGC 2682, making use of the data from \cite{1993A&AS..100..243Z}. The membership analyses were carried out utilising N = 2 and 4 variables, and they were compared with those obtained by the methodologies MT2, MT4 and our classification carried out using the membership probabilities present in the catalogue of \cite{1993A&AS..100..243Z}. The results obtained by our new methodology show a high degree of agreement with those obtained by the other membership analyses. Making use of the parameters of the distribution functions of both populations obtained for \cite{1993A&AS..100..243Z}, a CD of 1.57 has been calculated. For this value, and using the results obtained in our simulations, we estimate values of \textit{C} around 98\% for the methodologies that use N = 2 and 4 variables. The \textit{M} for this distance is around 5\%, that is, approximately 52 stars.

It is worth mentioning that in the present work we show the potential of our methodology in determining cluster members. The comparison with other widely used methodologies shows a high degree of agreement. However, unlike those methodologies, the new methodology presents a high flexibility in the use of different sets of variables. This feature enables us to carry out membership analyses adapting its application to the best variables available in each survey. We should not forget that our ultimate goal is Gaia, for which this code has been specially designed.

\section*{Acknowledgments}

We thank the referee for his/her comments and suggestions, which have heightened the quality of this work. We acknowledge Nestor Sanchez for sending the results of the membership analysis of NGC 2682. We acknowledge the IAA-CSIC for hosting L.S. during the time this paper was worked out. We acknowledge support from the Spanish Ministry for Economy and Competitiveness and FEDER funds through grant AYA2013-40611-P.

\label{lastpage}

\end{document}